\documentclass[onefignum,onetabnum]{siamart190516}

\usepackage{lipsum}
\usepackage{amsfonts}
\usepackage{algorithmic}
\usepackage{amssymb}
\usepackage{amsmath}
\usepackage{graphicx}
\usepackage{subfig}
\usepackage{epstopdf}
\usepackage{mathtools}
\usepackage{cleveref}
\usepackage{bm}

\DeclareMathOperator{\Tr}{Tr}

\ifpdf
\DeclareGraphicsExtensions{.eps,.pdf,.png,.jpg}
\else
\DeclareGraphicsExtensions{.eps}
\fi

\newsiamremark{remark}{Remark}
\newsiamremark{hypothesis}{Hypothesis}
\crefname{hypothesis}{Hypothesis}{Hypotheses}
\newsiamthm{claim}{Claim}

\headers{Eulearian nonlinear model for tissue growth}{C. Wei, M. Wu}

\title{An Eulerian nonlinear elastic model for compressible and fluidic tissue with radially symmetric growth
}

\author{Chaozhen Wei\footnotemark[2] 
\and Min Wu\footnotemark[2] 
}

\usepackage{amsopn}

\begin{document}

\maketitle
  
\renewcommand{\thefootnote}{\fnsymbol{footnote}}

\footnotetext[2]{Department of Mathematical Sciences, Worcester Polytechnic Institute, Worcester, MA 10605 USA
	(C.W.\email{cwei@wpi.edu}, M.W.\email{englier@gmail.com})}

\renewcommand{\thefootnote}{\arabic{footnote}}

\begin{abstract}
Cell proliferation, apoptosis, and myosin-dependent contraction can generate elastic stress and strain in living tissues, which may be dissipated by internal rearrangement through cell topological transition and cytoskeletal reorganization. Moreover, cells and tissues can change their sizes in response to mechanical cues. The present work demonstrates the role of tissue compressibility and internal rearranging activities on its size and mechanics regulation in the context of differential growth induced by a field of growth-promoting chemical factors. We develop a mathematical model based on finite elasticity and growth theory and the reference map techniques to describe the coupled tissue growth and mechanics in the Eulerian frame. We incorporate the tissue rearrangement by introducing a rearranging rate to the reference map evolution, leading to elastic-energy dissipation when tissue growth and deformation are in radial symmetry. By linearizing the model, we show that the stress follows the Maxwell-type viscoelastic relaxation. The rearrangement rate, which we call tissue fluidity, sets the stress relaxation time, and the ratio between the shear modulus and the fluidity sets the tissue viscosity. By nonlinear simulation of growing tissue spheroids and discs with graded growth rates along the radius, we find that the tissue compressibility and fluidity influence their equilibrium size. By comparing the nonlinear simulations with the linear analytical solutions, we show the size change as a nonlinear effect due to the advection of the tissue density flow, which only occurs when both tissue compressibility and fluidity are small. We apply the model to study tumor spheroid growth and epithelial disc growth when a reaction-diffusion process determines the growth-promoting factor field.
\end{abstract}



\section{Introduction}
\label{sec:introduction} 
The interaction between cell activities and tissue mechanics plays an essential role during development, homeostasis, and cancer progression. Mechanical stresses can arise and accumulate in growing tissues due to inhomogeneous cell proliferation and apoptosis, and can be dissipated by rearranging activities such as cell neighbor exchange, oriented cell divisions and cell extrusion \cite{lecuit2011force,Guillot-2013-extrusion,legoff2016mechanical}. Not only do cell activities generate mechanical forces, tissue growth also responds to local stress and strain, such as in growing tumor spheroids \cite{Helmlinger1997,montel2011stress,Delarue-2014-inhibition} and {\it Drosophila} wing discs \cite{Aegerter-Wilmsen-2007-regulation,legoff2013global}. In the light of increasing evidence and attention to the interaction between growth and mechanics, the present study seeks to improve our analytical understanding in the role of tissue mechanical properties on its size and mechanics regulation. 

Over the last three decades, the theory of finite elasticity and growth \cite{Rodriguez-1994-decomposition}, also called morphoelasticity \cite{goriely2017mathematics}, has become a powerful tool to understand tissue growth and mechanics, especially for soft tissues \cite{fung2013biomechanics}.  The heart of the theory is to decompose the deformation gradient into a growth stretch tensor followed by an elastic deformation tensor. The growth tensor captures the local volumetric growth due to new material production, while the elastic deformation tensor captures the mechanical response of the tissue continuum to maintain its integrity and results in residual stress. Within this framework, specific models have been developed to understand growth and morphogenesis in various scenarios, such as the formation of gut lumen ridges \cite{Li-2011-jmps,shyer2013villification,amar2013anisotropic} and brain convolutions \cite{Bayly-2013-cortical,tallinen2016growth}, tumor growth \cite{Goriely-2005-prl} and pathological cardiac growth \cite{Goktepe-2010-cardiac}, the re-epithelialization of the adult skin wound \cite{Wu-2015-wound}, etc. The interested reader is referred to the recent reviews \cite{Jones-2012-siam-review,Ambrosi-2019-growth-review} and book \cite{goriely2017mathematics} for more examples.

Since cells undergo growth, division, and apoptosis in response to their inherited drives and environmental cues, tissue growth depends on both the material (inherited) and the spatial (environmental) coordinates. Most previous finite elasticity and growth models consider the dependence of growth on the material coordinates, except in \cite{goriely2005differential} to the best of our knowledge. The Lagrangian description makes it difficult to study the interaction between tissue growth and environmental cues, such as the direct coupling between the tissue growth rate and the reaction-diffusion process of the growth-promoting factors (e.g., \cite{Cristini2003tumor}). 

Recently, one of us has developed an Eulerian model of incompressible tissue growth and elasticity \cite{Wu-2019-stress} by using the reference map techniques \cite{Cottet-2008-referencemap,Kamrin-2012-referencemap}. The reference map techniques reconstruct the deformation gradient based on the reference map, the inverse of the motion, instead of describing the dynamics of deformation gradient in the Eulerian frame \cite{Liu2001ARMA,Lin2005cpam}. In our model \cite{Wu-2019-stress}, we have modified the evolution equation of the reference map with a rearrangement rate to account for the tissue rearranging activities, which results in a Maxwell-type stress relaxation. We have applied this model to understand the regulation of tumor spheroid size and mechanics through diffusive growth-promoting factors and growth-inhibiting mechanical feedbacks, in which the steady state size of the tumor decreases when the strength of external physical confinement \cite{Helmlinger1997} and compression \cite{montel2011stress,Delarue-2014-inhibition} are elevated. In addition,  we have considered the tumor volume loss in response to mechanical compression \cite{Delarue-2014-inhibition} as an active process due to cell water efflux \cite{guo2017cell}. Alternatively, the cell and tissue can be considered a compressible material with measurable bulk moduli  \cite{guo2017cell,nolan2016compressibility}, which can change volume and density passively to local stress. 

In this paper, we develop an Eulerian model of tissue growth and elasticity considering its compressibility and rearranging activities. We assume that the soft tissues are compressible neo-Hookean materials with an isotropic growth rate in response to spatial stimuli and a rearrangement rate which changes their initial reference positions. With radially symmetric growth, we show that the rearranging activities in our formulation dissipate the stored elastic energy and enables the model to behave like a linear Maxwell viscoelastic model when the rearrangement rate is larger than the growth rate. By presenting linear analysis and nonlinear simulations for our model, we demonstrate the synergic effect of tissue compressibility and internal rearrangement on tissue size and mechanics regulation, and apply the model to study tumor spheroid growth and epithelial disc growth when the growth-promoting factor field is determined by a reaction-diffusion process.

This paper is organized as follows. In Sec.~\ref{sec:growing_tissues}, we present our Eulerian model with compressible tissue growth and elasticity using reference map techniques. In Sec.~\ref{sec:tissue-rearrange}, we introduce the rearranging activity in the reference map evolution equation in radial symmetry and derive the evolution equation of the elastic deformation and the strain energy. In Sec.~\ref{sec:linear_model}, we linearize the model when the time scale of rearranging activity is much shorter than that of the growth in radial symmetry. We show the linearized model behaves like the Maxwell viscoelastic model at the time scale of rearrangement rate and like viscous fluid at the time scale of growth. In Sec.~\ref{sec:prescribed_growth}, we show the synergic effect of tissue compressibility and  rearrangement on tissue size and mechanics by presenting the steady-state analytical solution for the linearized model and performing numerical simulations for nonlinear model. In Sec.~\ref{sec:coupling_growth}, we apply our nonlinear model to investigate tumor growth where the growth rate is coupled with a growth-promoting factor field determined by a reaction-diffusion process. In Sec.~\ref{sec:discussion}, we summarize and discuss our results.



\section{Growing Tissues as Compressible Elastic Material}\label{sec:growing_tissues}
\subsection{Decomposition of Deformation Gradient}\label{subsec:neo-hookean}
We consider a region of soft tissues initially in a stress-free reference configuration $\mathcal{B}_0\subset\mathbb{R}^d$ ($d=2$ or $3$). The mapping $\boldsymbol\chi:\mathcal{B}_0\rightarrow\mathcal{B}_t$ is introduced to relate the stress-free initial reference configuration $\mathcal{B}_0$ to the current deformed configuration $\mathcal{B}_t$ via $\mathbf{x}=\boldsymbol\chi(\mathbf{X},t)$, where $\mathbf{x} \in \mathcal{B}_t$ is the current location at time $t$ of a particle initially located at $\mathbf{X}\in\mathcal{B}_0$. The tangent map of $\boldsymbol\chi$ is denoted by $\mathbf{F}=\partial \mathbf{x}/\partial \mathbf{X}$, the {\it geometric deformation gradient}.

\begin{figure}[!b]
	\centering
	\includegraphics[width=1.0\textwidth]{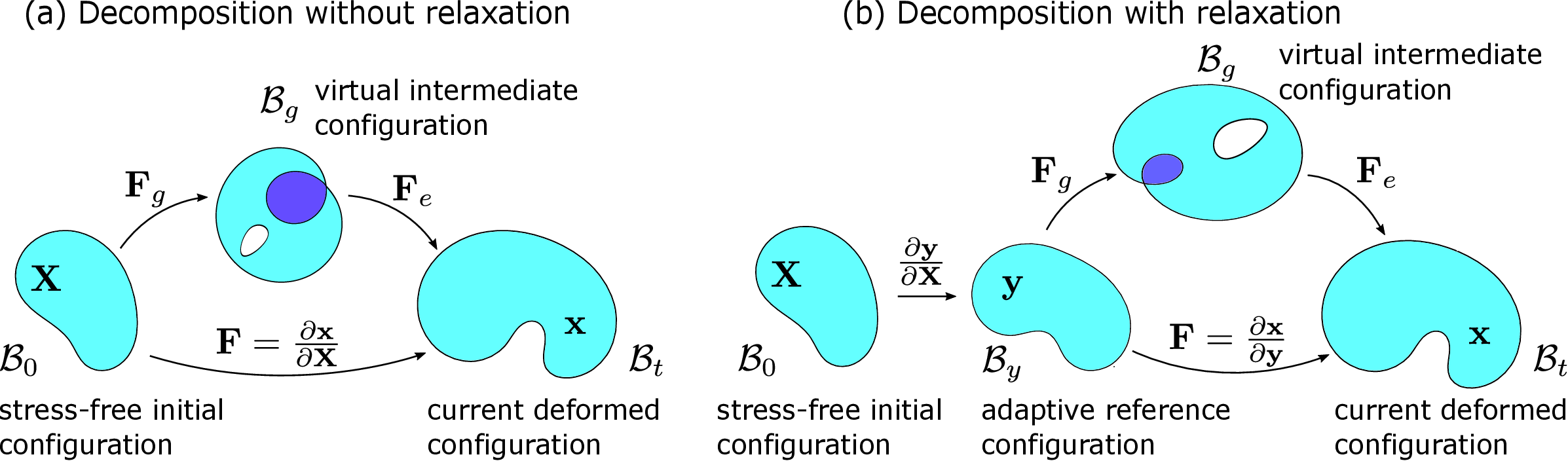}
	\caption{Decomposition of the geometric deformation gradient into growth stretch tensor and elastic deformation tensor. (a) Loop $\mathcal{B}_0\rightarrow\mathcal{B}_g\rightarrow\mathcal{B}_t$: decomposition of the geometric deformation with respect to stress-free initial configuration. (b) loop $\mathcal{B}_0\rightarrow\mathcal{B}_y\rightarrow\mathcal{B}_g\rightarrow\mathcal{B}_t$: relaxation of initial configuration to adaptive reference configuration and the decomposition of the geometric deformation with respect to adaptive reference configuration. }\label{fig:decomposition_map}
\end{figure}

In the theory of finite elasticity and growth \cite{Rodriguez-1994-decomposition}, we can decompose the geometric deformation gradient by
\begin{equation}\label{eq:decomp}
\mathbf{F}= \mathbf{F}_e \mathbf{F}_g
\end{equation}
where $\mathbf{F}_g$ is the accumulated {\it growth stretch tensor} describing the stretch relative to the reference configuration due to stress-free volumetric tissue growth, which may result in an intermediate incompatible configuration $\mathcal{B}_g \subset R^d$. $\mathbf{F}_g$ can be interpreted as the tangent of the mapping from $\mathcal{B}_0$ to $\mathcal{B}_g\subset R^D$ ($D>d$) when $\mathcal{B}_g$ is a differential manifold. The {\it elastic deformation tensor} $\mathbf{F}_e$ describes the elastic response of the tissue material and maps the tangent space of $\mathcal{B}_g$ to that of $\mathcal{B}_t$. \Cref{fig:decomposition_map}(a) illustrates this decomposition of geometric deformation. The idea was developed much earlier in research communities other than biomechanics \cite{sadik2017origins} and is frequently called the Kr\"oner-Lee decomposition in elastoplasticity \cite{kroner1959allgemeine,lee1969elastic}.

\subsection{Dynamics of Reference Map and Deformation Gradient}
Instead of describing the geometric deformation map $\boldsymbol\chi(\mathbf{X},t)$ in the Lagrangian frame, we introduce a system of finite elasticity and growth in the Eulerian frame by using the {\it reference map} \cite{Cottet-2008-referencemap,Kamrin-2012-referencemap}, $\mathbf{X}=\mathbf{Y}(\mathbf{x},t):=\boldsymbol\chi^{-1}(\mathbf x,t)$, which is the inverse of the motion $\boldsymbol\chi$. Since the initial reference coordinates of a material point does not change in time, we have $d\mathbf{X}/dt=0$, where $d f/dt = \partial f(\mathbf{X},t)/\partial t$ is the material time derivative of any function $f$. Letting $\mathbf{v}(\mathbf{x},t)=\frac{\partial}{\partial t}\boldsymbol\chi(\mathbf{X},t)$ be the material velocity, and using the  chain rule, we obtain the dynamics of reference map (with the initial condition)
\begin{equation}\label{eq:y_evol1}
\frac{d{\mathbf{X}}}{dt}=\frac{\partial \mathbf{Y}}{\partial t}(\mathbf{x},t)+\mathbf{v}\cdot\nabla\mathbf{Y}(\mathbf{x},t)=\mathbf{0},~~\text{with}~~ \mathbf{Y}(\mathbf{x},0)=\mathbf{x}.
\end{equation}

We can show that Eq.~\cref{eq:y_evol1} leads to the dynamics of the deformation gradient $\mathbf{F}=\partial \mathbf{x}/\partial \mathbf{X}=(\partial \mathbf{X}/\partial \mathbf{x})^{-1}=(\nabla\mathbf Y)^{-1}$ (where $\nabla=\nabla_{\mathbf{x}}$) \cite{Cottet-2008-referencemap,Kamrin-2012-referencemap}. By taking $\nabla$ on both sides of Eq.~\cref{eq:y_evol1}, we obtain the evolution of $
d(\nabla\mathbf{Y})/dt={\partial (\nabla\mathbf Y)}/{\partial{{t}}}+ {\mathbf v} \cdot {\nabla} {(\nabla\mathbf Y)}=-\nabla\mathbf Y \nabla\mathbf{v}$. Together with the relation $d(\nabla\mathbf{Y}^{-1})/dt =- \nabla\mathbf{Y}^{-1}(d(\nabla\mathbf{Y})/dt)\nabla\mathbf{Y}^{-1}$, we obtain the evolution of $(\nabla\mathbf{Y})^{-1}$
\begin{equation}\label{eq:F_evol1}
\frac{d (\nabla\mathbf{Y})^{-1}}{d{{t}}}=\frac{\partial (\nabla\mathbf{Y})^{-1}}{\partial{{t}}}+ {\mathbf v} \cdot {\nabla} {(\nabla\mathbf{Y})^{-1}}= \nabla\mathbf{v}(\nabla\mathbf{Y})^{-1},~~\text{with}~~(\nabla\mathbf{Y})^{-1}(\mathbf{x},0)=\mathbf{I}
\end{equation}
which is equivalent to the Lagrangian formalism ${d\mathbf{F}}/{dt}=\nabla\mathbf{v}\mathbf{F}$ with ${\mathbf{F}(\mathbf{X},t)=\mathbf{I}}$. Following this, we can obtain the evolution of the local volumetric (when $d=3$) or areal (when $d=2$) variation  $J=\det(\mathbf{F})=\det(\nabla \mathbf Y)^{-1}$. Using the chain rule $dJ/dt=[\partial J/\partial (\nabla\mathbf{Y})^{-1}]:[d(\nabla\mathbf{Y})^{-1}/dt]$ (where $A:B=A_{ij}B_{ij}$ is the Frobenius inner product of two matrix represented in Einstein notation) and the relation $\partial J/\partial(\nabla\mathbf{Y})^{-1}=J(\nabla\mathbf{Y})^{\text{T}}$, we obtain 
\begin{equation}\label{eq:J_evol1}
\frac{dJ}{dt} =\frac{\partial J}{\partial{{t}}}+ {\mathbf v} \cdot {\nabla} {J}= J\nabla\cdot\mathbf{v},~~\text{with}~~J(\mathbf{x},0)=1
\end{equation}
where $\nabla \cdot \mathbf{v}$ represents the rate of local volumetric (when $d=3$) or areal (when $d=2$) expansion. The above derivations, which lead to Eqs.\cref{eq:F_evol1,eq:J_evol1}, demonstrate that the dynamics of $\mathbf{Y}$ in Eq.~(\ref{eq:y_evol1}) implies the dynamics of $\mathbf{F}=(\nabla\mathbf{Y})^{-1}$ and $J=\det \mathbf{F}$. Thus, we only need to trace the evolution of $\mathbf{Y}$ \cref{eq:y_evol1}, rather than explicitly solve Eqs.~\cref{eq:F_evol1,eq:J_evol1} in this Eulerian formulation.

\subsection{Dynamics of Active Growth}
According to the decomposition Eq.~\cref{eq:decomp}, the local volume variation $J=J_gJ_e$,  where $J_g=\det(\mathbf{F}_g)$, {\it the active volume variation} due to growth and $J_e=\det(\mathbf{F}_e)$,  {\it the elastic volume variation} due to elastic deformation. Defining the current tissue density $\rho(\mathbf{x},t)$ (mass per current unit volume), by change of variable, we have
\begin{equation}\label{eq:dv_balance}
\int_{\Omega_t}\rho dV_t = \int_{\Omega_0}\rho J dV_0,
\end{equation}
where $\Omega_t\subset\mathcal{B}_t$ denotes an arbitrary current region and coincides with $\Omega_0\subset\mathcal{B}_0$ at $t=0$, and $dV_t\subseteq\mathcal{B}_t$ ($dV_0\subseteq\mathcal{B}_0$) denotes the infinitesimal volume element in the current (initial stress-free) configuration. 

To describe the evolution of growth, we consider
\begin{equation}\label{eq:mass_rate_balance}
\frac{d}{dt}\int_{\Omega_t} \rho \ dV_t=\int_{\Omega_t} \rho\gamma \ dV_t,
\end{equation}
where $\gamma(\mathbf{x},t)$ is the rate of mass production or loss due to cell proliferation and apoptosis, which can depend on the spatial coordinates $\mathbf{x}$ via chemical and mechanical factor regulations \cite{Helmlinger1997,montel2011stress,Delarue-2014-inhibition,Aegerter-Wilmsen-2007-regulation,legoff2013global}. 
By Reynolds transport theorem and the arbitrariness of $\Omega_t$, we obtain the equation of mass balance with source
\begin{equation}\label{eq:rho_evol1}
\frac{d\rho}{dt} =\frac{\partial \rho}{\partial t}+\mathbf{v}\cdot\nabla\rho= \rho(\gamma - \nabla\cdot \mathbf{v}), ~~\text{with}~~\rho(\mathbf{x},0)=\rho_0,
\end{equation}
where $\rho_0$ is the initial stress-free tissue density.

To consider mass balance between the stress-free configuration and the current configuration, we assume that the stress-free density $\rho_0$ remains constant during growth \cite{Jones-2012-siam-review}. That is, the change of mass is only attributed to the active volume variation $J_g$; the elastic volume variation $J_e$ only changes the local density and volume element but not the mass. After active growth, the volume element $dV_0$ is modified into $J_gdV_0$, and $\rho$ and $\rho_0$ are connected by
\begin{equation}\label{eq:dm_balance}
\int_{\Omega_t} \rho dV_t = \int_{\Omega_0} \rho_0 (J_g dV_0).
\end{equation} 

Combining Eqs.~\cref{eq:dv_balance,eq:dm_balance},  we obtain the relation between densities 
\begin{equation}\label{eq:rho_balance}
\rho J = \rho_0 J_g
\end{equation}
due to the arbitrariness of $\Omega_0$. Together with the relation $J=J_gJ_e$, we have $J_e = \rho_0/\rho$.

By the assumption of constant $\rho_0$ and combining Eqs.~\cref{eq:J_evol1,eq:rho_balance,eq:rho_evol1}, we obtain the evolution of the  active volume variation
\begin{equation}\label{eq:Jg_evol1}
\frac{dJ_g}{dt}=\frac{\partial J_g}{\partial t}+\mathbf{v}\cdot\nabla J_g=\gamma J_g, ~~\text{with}~~J_g(\mathbf{x},0)=1.
\end{equation}
One can see that under the assumption that  $\rho_0$ remains constant during growth, the rates of mass production and active volumetric growth (when $d=3$) or areal growth (when $d=2$) are identical, both being $\gamma$. For simplicity, we assume the growth stretch tensor $\mathbf{F}_g = J_g^{1/d}\mathbf I$ to be isotropic such that its dynamics is determined by Eq.~\cref{eq:Jg_evol1}. We define the isotropic growth rate tensor
\begin{equation}
\boldsymbol{\Gamma}_{iso}:=\frac{d\mathbf{F}_g}{dt}\mathbf{F}_g^{-1}=\frac{\gamma}{d}\mathbf{I}.
\end{equation}

\subsection{Dynamics of Elastic Deformation}
By the relation $\mathbf{F}_e=\mathbf{F} \mathbf{F}_g^{-1}=J_g^{-1/d}\nabla \mathbf Y^{-1}$ and Eqs.~\cref{eq:F_evol1,eq:Jg_evol1}, we obtain the evolution of $\mathbf{F}_e$
\begin{equation}\label{eq:F_e_evol1}
\frac{d \mathbf{F}_e}{dt}=\frac{\partial \mathbf{F}_e}{\partial t} + \mathbf{v}\cdot \nabla \mathbf{F}_e = (\nabla\mathbf{v}-\boldsymbol{\Gamma}_{iso})\mathbf{F}_e, ~~\text{with}~~\mathbf{F}_e(\mathbf{x},0)=\mathbf{I}
\end{equation}
which yields the dynamics of the elastic volume (when $d=3$) or areal (when $d=2$) variation
\begin{equation}\label{eq:Je_evol1}
\frac{dJ_e}{dt}=\frac{\partial J_e}{\partial t}+\mathbf{v}\cdot\nabla J_e=J_e(\nabla\cdot\mathbf{v}-\gamma), ~~\text{with}~~J_e(\mathbf{x},0)=1.
\end{equation}
The evolution of $J_e$ is consistent with the relation $J_e = \rho_0/\rho$ and Eq.~\cref{eq:rho_evol1}. The above derivations demonstrate that the dynamics of $\mathbf{Y}$ and $J_g$ in Eqs. \cref{eq:y_evol1,eq:Jg_evol1} imply the dynamics of $\mathbf{F}_e=J_g^{-1/d}(\nabla\mathbf{Y})^{-1}$,  $J_e=\det \mathbf{F}_e$, and $\rho=\rho_0/J_e$. Thus, mass balance and elastic deformation dynamics are ensured by solving Eqs. \cref{eq:y_evol1,eq:Jg_evol1} in this Eulerian formulation.


\subsection{Compressible neo-Hookean Elasticity}
For simplicity, we treat soft tissues as isotropic, compressible neo-Hookean elastic material. The strain energy density function (per unit stress-free volume) in $d$-dimensions is given by \cite{Taber2009} \begin{equation}\label{eq:strain_enegy_density}
W(\mathbf{F}_e)=\frac{1}{2}\mu(\bar{I}_1-d)+\frac{1}{2}K(J_e-1)^2,
\end{equation}
where $\mu$ and $K$ are, respectively, the shear and bulk modulus, and $\bar{I}_1=J_e^{-2/d}\Tr(\mathbf{C}_e)$ is the first invariant of the isochoric part $\bar {\mathbf{C}}_e=J_e^{-1/d}\mathbf{C}_e$ of the {\it right Cauchy–Green elastic deformation tensor} $\mathbf{C}_e=\mathbf{F}_e^\text{T}\mathbf{F}_e$.

To derive the stress-strain constitutive relation for compressible elastic body with growth, we consider the total strain energy of a region $\Omega_t\subseteq\mathcal{B}_t$
\begin{equation}\label{eq:strain_energy_tot}
E(t)=\int_{\Omega_t} J_e^{-1}W(\mathbf{F}_e) dV_t,
\end{equation}
where the factor $J_e^{-1}$ accounts for the ratio of the stress-free volume element after growth $J_gdV_0$ to the deformed volume element $dV=JdV_0$. Taking the material time derivative of $E(t)$ leads to 
\begin{align}\nonumber
\frac{d E}{dt} &= \int_{\Omega_t} \Big(\frac{d}{dt}(J_e^{-1}W) + J_e^{-1}W \nabla \cdot \mathbf{v} \Big) dV_t, \\ \nonumber  
&= \int_{\Omega_t} \Big(J_e^{-1}\frac{\partial W}{\partial \mathbf{F}_e}:\frac{d\mathbf{F}_e}{dt} -J_e^{-2}W\frac{dJ_e}{dt} + J_e^{-1}W \nabla \cdot \mathbf{v} \Big) dV_t, \\ \label{eq:dE}
&= \int_{\Omega_t} \Big(J_e^{-1}\frac{\partial W}{\partial \mathbf{F}_e}\mathbf{F}_e^{\text T}: \nabla \mathbf{v} - J_e^{-1}\frac{\partial W}{\partial \mathbf{F}_e}\mathbf{F}_e^{\text T}:\boldsymbol{\Gamma}_{iso} +  J_e^{-1}W\gamma\Big) dV_t,
\end{align}
where we have used the evolution equations \cref{eq:F_e_evol1,eq:Je_evol1} in the last line. We assume that the growth does not interfere the elasticity of tissue material. When $\gamma=0$ and $\mathbf\Gamma_{iso}=\mathbf{0}$, we obtain the constitutive relation for the {\it Cauchy stress tensor}
\begin{align} 
\boldsymbol\sigma &= {J_e}^{-1}\frac{\partial W}{\partial \mathbf{F}_e} \mathbf{F}_e^{\text T},\nonumber\\ \nonumber 
&=\mu J_e^{-\frac{2+d}{d}}\Big(\mathbf{F}_e\mathbf{F}_e^{\text{T}}-\frac{I_1}{d}\mathbf{I}\Big)+K(J_e-1)\mathbf{I}, \\ 
&= \mu J_g J^{-\frac{2+d}{d}}\Big(\nabla \mathbf{Y}^{-1}\nabla\mathbf{Y}^{-\text{T}}-\frac{\tilde{I}_1}{d}\mathbf{I}\Big)+K(J_g^{-1}J-1)\mathbf{I}, \label{eq:constitutive_law}
\end{align}
where $I_1=\Tr(\mathbf{C}_e)$ and $\tilde{I}_1=\Tr(\mathbf{C})$ where $\mathbf{C}=\nabla\mathbf{Y}^{-\text{T}}\nabla \mathbf{Y}^{-1}=\mathbf{F}^T\mathbf{F}$ is the {\it right Cauchy–Green deformation tensor}. The stress can be decomposed as $\boldsymbol{\sigma}=\boldsymbol{\sigma}_s+\boldsymbol{\sigma}_p$, the sum of the deviatoric part $\boldsymbol{\sigma}_s$ (proportional to $\mu$) and the isotropic part $\boldsymbol{\sigma}_p$ (proportional to $K$). For incompressible materials, i.e., $J_e\equiv 1$, the isotropic part of the stress vanishes. Instead, a pressure term should be introduced (as a Lagrangian multiplier) for maintaining incompressibility \cite{Wu-2019-stress}. 

\subsection{Dynamics of Growing Tissue Without Relaxation}
Recognizing the constitutive stress-strain relation, the rate of change of total strain energy reduces to 
\begin{equation}\label{eq:oldenergy}
\nonumber
\frac{dE}{dt}=\int_{\Omega_t} J_e^{-1}W\gamma dV_t - \int_{\Omega_t} \boldsymbol{\sigma}_p:\boldsymbol{\Gamma}_{iso} dV_t -\int_{\Omega_t} (\nabla\cdot \boldsymbol{\sigma})\cdot\mathbf{v} dV_t + \int_{\partial \Omega_t} \mathbf{v}\cdot (\boldsymbol{\sigma}\cdot \mathbf{n}) dS_t,
\end{equation}
where $\mathbf{n}$ is the outward unit normal of the boundary $\partial \Omega_t$ and we have used the relations $\boldsymbol{\sigma}:\boldsymbol{\Gamma}_{iso}=\boldsymbol{\sigma}_p:\boldsymbol{\Gamma}_{iso}$, $\boldsymbol{\sigma}:\nabla \mathbf{v} = \nabla\cdot(\boldsymbol{\sigma}\cdot\mathbf{v}) - (\nabla\cdot \boldsymbol{\sigma})\cdot\mathbf{v}$ and the divergence theorem. The first term on the right-hand side represents the change of elastic energy due to the addition of material by volumetric (or areal) growth. Without the minus sign, the second term represents the stress power due to active isotropic growth, similar to the stress power due to local geometric volumetric (or areal) expansion, except that $\boldsymbol{\sigma}_p$ is conjugated with isotropic growth rate tensor $\boldsymbol{\Gamma}_{iso}$ instead of local volumetric (or areal) expansion $(\nabla\cdot \mathbf{v})\mathbf{I}$. This shows that the active growth ($\gamma>0$) in the compressed tissue ($\Tr(\boldsymbol{\sigma}_p)<0$) requires the addition of energy. The third and fourth term represent the stress power (due to the motion) in the bulk and at boundary. In the theory of finite elasticity and growth \cite{Rodriguez-1994-decomposition}, the body force and inertial force are often assumed to be negligible.  Following this assumption, we obtain the equation of mechanical equilibrium
\begin{equation}\label{eq:equil_eqn}
\nabla \cdot \boldsymbol{\sigma} = \mathbf 0
\end{equation}
in the tissue bulk $\Omega_t$. At the fixed boundary $\Sigma_D$, we prescribe
\begin{equation}\label{eq:bc0}
\mathbf{v} = \mathbf0,
\end{equation}
and at the free boundary $\Sigma_t$, we prescribe
\begin{equation}\label{eq:bc1}
\boldsymbol{\sigma}\cdot \mathbf{n} = \mathbf{0}.
\end{equation}

Combining the evolution equations \cref{eq:y_evol1,eq:Jg_evol1} with their initial conditions, the constitutive relation (\ref{eq:constitutive_law}), the mechanical equilibrium equation (\ref{eq:equil_eqn}), and the boundary condition equations (\ref{eq:bc0}) and (\ref{eq:bc1}), we establish a coupled nonlinear system for describing the tissue growth and mechanics without stress relaxation.

\section{Tissue Rearrangement and Stress Relaxation}\label{sec:tissue-rearrange}
At the large-scale continuum level, soft tissues are effectively viscoelastic due to cell neighbor exchanges  \cite{lecuit2011force,legoff2016mechanical} and cytoskeleton reorganization \cite{doubrovinski2017measurement} at the cell and sub-cellular level. Therefore, the stress caused by differential growth can be effectively relaxed; otherwise, the stress may accumulate and blow up. In the theory of finite elasticity and growth, stress-dependent growth rate can be used to drive the stress relaxation to a specified target stress \cite{Taber2009}. One could also extend the traditional theory by introducing viscoelastic mechanisms to relax elastic strains (the Kelvin-Voigt model) or stresses (the Maxwell model) \cite{Lubkin2002,Bresch2009Spectial}.  In addition, anisotropic growth due to orientated cell divisions and apoptosis may effectively alleviate the accumulated stress and introduce ``fluidity" in the tissue mechanics \cite{araujo2005mixture,Ranft-2010-pnas}. Following \cite{Wu-2019-stress}, we will consider the stress relaxation mechanism for compressible tissues by modifying the reference map evolution. The new development in this section is to derive the evolution equation for the ``adaptive" active volume variation $J_g$, which is implicitly satisfied in the incompressible case (see below) and prove that our relaxation mechanism dissipates the stored strain energy where rotation can be neglected.

\subsection{Adaptive Reference Map}\label{subsec:adaptive-reference-map} We describe the tissue rearranging activities by introducing an ``adaptive" reference configuration $\mathcal{B}_y$ that relaxes from the initial reference configuration $\mathcal{B}_0$ and adapts to the current deformed configuration $\mathcal{B}_t$. We describe the ``relaxed" reference coordinates by the adaptive reference map $\mathbf{Y}:\mathcal{B}_t\rightarrow\mathcal{B}_y$ via $\mathbf{y}=\mathbf{Y}(\mathbf{x},t)$ for a particle currently located at $\mathbf{x}\in \mathcal{B}_t$ and initially located at $\mathbf{X}\in\mathcal{B}_0$.
Henceforth we will use the same notations as in Sec.~\ref{sec:growing_tissues} to represent all tensors and their determinant in the presence of tissue rearrangement (i.e., with respect to the adaptive reference configuration). 
The relation of the adaptive reference $\mathcal{B}_y$ with $\mathcal{B}_0$ and $\mathcal{B}_t$ is illustrated in Fig.~\ref{fig:decomposition_map}(b). 

Now we can define the effective deformation gradient $\mathbf{F}=\partial \mathbf{x}/\partial \mathbf{y}=\nabla \mathbf{Y}^{-1}$, respect to the adaptive reference configuration $\mathcal{B}_y$. In general, $\mathbf{F}=\mathbf{R}\mathbf{U}$ by the polar decomposition, where $\mathbf{U}=\sqrt{\mathbf{F}^T\mathbf{F}}=\sqrt{\mathbf{C}}$ is the symmetric positive-definite stretch tensor and  $\mathbf{R}=\mathbf{F}\mathbf{U}^{-1}$ is the orthogonal rotation tensor. For the symmetric tissue growth, such as tumor spheroid growth or circular wound closure processes, $\mathbf{R}=\mathbf{I}$. Henceforth, for simplicity, we will focus our discussion on the tissue growth with rearrangement where rotation can be neglected, and therefore we will write the deformation gradient $\mathbf{F}=\nabla \mathbf{Y}^{-1}=\mathbf{U}$ and the elastic deformation tensor $\mathbf{F}_e=\sqrt{\mathbf{C}_e}=\mathbf{U}_e$, the elastic stretch tensor. The system with rotation ($\mathbf{R}\neq\mathbf{I}$) is discussed in Supplementary Materials.

\subsection{Dynamics with Tissue Rearrangement} 
The dynamics of the adaptive reference map $\mathbf{y}=\mathbf{Y}(\mathbf{x},t)$ in Eulerian frame is given by
\begin{equation}\label{eq:y_evol2}
\frac{d\mathbf{y}}{dt}=\frac{\partial \mathbf{Y}}{\partial t}+\mathbf{v}\cdot\nabla \mathbf{Y}=\beta(\mathbf x-\mathbf{y}),~~\text{with}~~ \mathbf{Y}(\mathbf{x},0)=\mathbf{x}
\end{equation}
where $\beta$ is the rate of rearrangement that measures how fast the reference configuration of the tissue adapts to the current configuration. In principle, $\beta$ can vary spatiotemporally, regulated by oriented cell divisions that distribute new cells in favor of relaxing stresses, or without division, by cell-cell and cytoskeleton rearrangements. Here we consider it as spatially uniform for simplicity. When $\beta\neq0$, this evolution equation can be rewritten as the relaxation of $\mathbf{u}=\mathbf{x}-\mathbf{y}$ with decaying rate $\beta$, $d\mathbf{u}/dt=\mathbf{v}-\beta\mathbf{u}$. We call $\mathbf{u}$ the {\it adaptive displacement} henceforth, since $\mathbf u$ does not follow the original relation between  displacement and velocity $d\mathbf{u}/dt=\mathbf{v}$ when $\beta\neq0$.  When $\beta=0$, we recover the dynamics of the original reference map without tissue arrangement in Eq.~(\ref{eq:y_evol1}) and the relation $d\mathbf{u}/dt=\mathbf{v}$.


Following similar derivation process in Sec.~\ref{sec:growing_tissues}, the dynamics of adaptive reference map (\ref{eq:y_evol2}) leads to the dynamics of the relaxed deformation gradient $\nabla \mathbf{Y}^{-1}=\mathbf{U}$
\begin{equation}\label{eq:F_evol2}
\frac{\partial \mathbf U}{\partial{{t}}}+ {\mathbf v} \cdot {\nabla} {\mathbf U}= [\nabla\mathbf{v}-\beta (\mathbf U-\mathbf{I})]\mathbf U, ~~\text{with}~~ \mathbf{U}(\mathbf{x},0)=\mathbf{I}.
\end{equation}
The term with $\beta$ demonstrates the relaxation of deformation gradient in response to the {\it Biot strain tensor} $(\mathbf U-\mathbf{I})$. This equation further leads to the evolution of the relaxed volume variation $J=\det{\mathbf U}$
\begin{equation}\label{eq:J_evol2}
\frac{dJ}{dt} =\frac{\partial J}{\partial t}+\mathbf{v}\cdot\nabla J= J\big[\nabla\cdot\mathbf{v} -\beta\Tr(\mathbf U-\mathbf{I})\big], ~~\text{with}~~ J(\mathbf{x},0)=1.
\end{equation}

We assume that the tissue rearranging activities does not interfere the dynamics of tissue density $\rho$ or the initial stress-free density $\rho_0$ (mass is conserved with considering rearrangement). Combining Eqs.~\cref{eq:rho_balance,eq:rho_evol1} with the adaptive evolution of $J$, Eq.~(\ref{eq:J_evol2}), the evolution of adaptive $J_g$ should follow
\begin{equation}\label{eq:Jg_evol2}
\frac{dJ_g}{dt} =\frac{\partial J_g}{\partial t}+\mathbf{v}\cdot\nabla J_g= J_g\big[\gamma -\beta\Tr(\mathbf U-\mathbf{I})\big], ~~\text{with}~~ J_g(\mathbf{x},0)=1.
\end{equation}
When strict incompressibility is imposed ($J_e=J/J_g=1$ in \cite{Wu-2019-stress}), comparing Eq.~\cref{eq:Jg_evol2} with \cref{eq:J_evol2} yields the constraint $\nabla\cdot\mathbf{v}=\gamma$. In fact,  Eq.~(\ref{eq:Jg_evol2}) is equivalent to $\nabla\cdot\mathbf{v}=\gamma$ when $J_e=J/J_g=1$ is given. Thus, Eq.~(\ref{eq:Jg_evol2}) describes the dynamics of $J_g$ in both compressible and incompressible rearranging tissues.

Assuming that the adaptive growth stretch tensor $\mathbf{F}_g=J_g^{1/d}\mathbf{I}$ is still isotropic, the dynamics of the adaptive elastic stretch tensor $\mathbf{U}_e=\mathbf{F}_e=J_g^{-1/d}\mathbf{U}$ follows
\begin{equation}\label{eq:F_e_evol2}
\frac{d\mathbf{U}_e}{dt}=\frac{\partial \mathbf{U}_e}{\partial t} + \mathbf{v}\cdot \nabla \mathbf{U}_e = [(\nabla\mathbf{v}-\boldsymbol{\Gamma}_{iso})-\beta(\mathbf U-\frac{1}{d}\Tr(\mathbf U)\mathbf{I})]\mathbf{U}_e \text{ with }\mathbf{U}_e(\mathbf{x},0)=\mathbf{I}.
\end{equation}
The term with $\beta$ is the simplification of  $[(\mathbf U-\mathbf{I})-\frac{1}{d}\Tr(\mathbf U-\mathbf{I})\mathbf{I}]$, showing the relaxation of $\mathbf{U}_e$ due to the deviatoric components of the Biot tensor. One can check that the volumetric change due to elastic deformation $J_e=\det(\mathbf{U}_e)$ yields the same dynamics $dJ_e/dt=J_e(\nabla\cdot\mathbf{v}-\gamma)$ as in Eq.~(\ref{eq:Je_evol1}) and the relation $\rho J_e = \rho_0$ still holds. 

Substituting Eqs.~(\ref{eq:Je_evol1}) and (\ref{eq:F_e_evol2}) into Eq.~(\ref{eq:dE}), we obtain the rate of change of the total strain energy 
\begin{equation}\nonumber
\frac{dE}{dt}=\int_{\Omega_t} \Big(J_e^{-1}W\gamma+ \boldsymbol{\sigma}:(\nabla \mathbf{v}-\boldsymbol{\Gamma}_{iso})\Big) dV_t -\beta \int_{\Omega_t} \boldsymbol{\sigma}_s:(\mathbf U-\frac{1}{d}\Tr(\mathbf U)\mathbf{I}) dV_t.
\end{equation}
The first integral represents the same energy change due to active growth as in Eq.~(\ref{eq:oldenergy}). The second integral with $\beta$ represents the energy change due to tissue rearrangement, where we have used the fact $\boldsymbol\sigma_p:(\mathbf U-\frac{1}{d}\Tr(\mathbf U)\mathbf{I})=0$ . From Eq.~(\ref{eq:constitutive_law}), one can check that the second integral is nonnegative and only vanishes when $\mathbf{U}$ is isotropic or $\boldsymbol{\sigma}_s=\mathbf{0}$. This means that the second integral with $\beta>0$ dissipates the stored strain energy only when there are both shear stress and shear deformation.


\section{Linearized Model Behavior}\label{sec:linear_model}
In this section, we will demonstrate the behavior of our nonlinear model with tissue rearrangement in its linearized regime. In the theory of small deformations, the adaptive displacement $\mathbf{u}=\mathbf{x}-\mathbf{y}$ is assumed to be small such that $\|\mathbf{u}\|\sim O(\epsilon)$ and $\|\nabla\mathbf{u}\|\sim O(\epsilon)$ are small where $\epsilon\ll 1$. Considering the characteristic time $T\sim 1/\beta$, we obtain the leading-order approximation of Eq.~(\ref{eq:y_evol2}) by the evolution of $\mathbf{u}$
\begin{equation}\label{eq:u_evol}
\frac{\partial \mathbf{u}}{\partial t} = \mathbf{v}-\beta\mathbf{u},~~~\text{with}~~\mathbf{u}(\mathbf{x},t)=\mathbf{0}\end{equation} 
where $\|\mathbf{v}\|\sim O(\epsilon)$ and the advection term is dropped as the higher order term of $\epsilon$.

By assuming that the rate of tissue material growth is slower than the rate of tissue rearrangement, i.e., $\gamma/\beta\sim O(\epsilon)$, the active volumetric or areal variation can be written as $J_g = 1+g$ where $g(\mathbf{x},t)\sim O(\epsilon)$ is the local active volume or areal increment. We can derive the evolution of $g$ by the leading-order approximation of Eq.~(\ref{eq:Jg_evol2}):
\begin{equation}\label{eq:g_evol}
\frac{\partial g}{\partial t} =  \gamma-\beta\nabla\cdot\mathbf{u},~~~\text{with}~~g(\mathbf{x},t)=0
\end{equation}
where again, the advection term is dropped as the higher order term of $\epsilon$.

The linearized stress from leading-order approximation of Eq.~(\ref{eq:constitutive_law}) can be written into the isotropic part and the deviatoric part  \begin{align}
\boldsymbol\sigma_p &= K(\nabla\cdot\mathbf{u}-g)\mathbf{I}, \label{eq:pressure_linear} \\ 
\boldsymbol\sigma_s &= \mu (\nabla \mathbf{u}+\nabla\mathbf{u}^{\text{T}}-(2/d)(\nabla\cdot\mathbf{u})\mathbf{I}). \label{eq:stress_linear}
\end{align}

\subsection{Viscoelastic Stress Relaxation}\label{subsec:multiscale_behavior} 
Differentiating Eq.~(\ref{eq:stress_linear}) in time and using Eq.~(\ref{eq:u_evol}), we obtain the Maxwell model of viscoelastic materials
\begin{equation}\label{eq:stress_s_relax}
\boldsymbol\sigma_s+\frac{1}{\beta}\frac{\partial {\boldsymbol\sigma}_s}{\partial t}=\frac{\mu}{\beta}\Big(\nabla \mathbf{v}+\nabla\mathbf{v}^{\text{T}}-\frac{2}{d}(\nabla\cdot\mathbf{v})\mathbf{I}\Big),~~~\text{with}~~\boldsymbol{\sigma}_s(\mathbf{x},t)=\mathbf{0}
\end{equation}
where $(\nabla \mathbf{v}+\nabla\mathbf{v}^{\text{T}})$ defines the strain rate, $\beta$ serves as the rate of stress relaxation and the ratio $\mu/\beta$ plays the role of ``effective'' viscosity. Note here we use the time derivative $\partial\boldsymbol\sigma_s/\partial t$ in the linear approximation. In the general case of large deformation, the upper-convected time derivative  $\stackrel{\triangledown}{\boldsymbol\sigma}= d\boldsymbol\sigma/dt - (\nabla \mathbf{v})^{\text{T}} \cdot \boldsymbol\sigma - \boldsymbol\sigma \cdot (\nabla \mathbf{v})$ should be present. The Maxwell-like relaxation of deviatoric stress \cref{eq:stress_s_relax} reflects the fluidization of tissue by rearrangement, and hence we can use the rate of relaxation $\beta$ to describe the fluidity of tissue. The tissue fluidization was also found by linear viscoelastic model with the stress relaxation timescale coupled with the orientated cell division and death \cite{Ranft-2010-pnas}. However, in general, the tissue rearrangement is not necessarily coupled with cell division but can be also facilitated by cell-cell topological transition and the reorganization of the cytoskeletal elements. In our linearized model, the tissue viscosity can be independently modulated by the rate of rearrangement $\beta$. Henceforth we call $\beta$ the {\it tissue fluidity}. 

Differentiating Eq.~(\ref{eq:pressure_linear}) in time and using Eq.~(\ref{eq:u_evol}), we obtain the leading-order evolution of the isotropic stress (or the pressure)
\begin{equation}\label{eq:stress_p_relax}
\frac{\partial {\boldsymbol\sigma}_p}{\partial t}=K(\nabla\cdot{\mathbf{v}}-\gamma)\mathbf{I}, ~~~\text{with}~~\boldsymbol{\sigma}_p(\mathbf{x},t)=\mathbf{0}
\end{equation}
where $K$ represents the bulk modulus. Clearly $\gamma$ can influence ${\partial {\boldsymbol\sigma}_p}/{\partial t}$, which is consistent with the pressure dynamics description in \cite{Ranft-2010-pnas}. Although the dynamics of $\boldsymbol{\sigma}_p$ does not include a dissipation term explicitly, the dissipation of $\boldsymbol{\sigma}_s$ through Eq.~(\ref{eq:stress_s_relax}) can influence $\boldsymbol{\sigma}_p$ through force balance equation $\nabla\cdot (\boldsymbol{\sigma}_s+\boldsymbol{\sigma}_p)=0$, which means the rate of rearrangement $\beta$ can influence both the pressure and deviatoric stress. This is different from \cite{Ranft-2010-pnas}, where an independent relaxation time of pressure is introduced to describe the tissue mechanics close to homeostasis.

\subsection{Viscous Tissue Flow in the Timescale of Growth}
If we assume the characteristic time  $T\sim 1/\gamma\gg1/\beta$, we have the quasi-steady-state relations from the leading order of Eqs.~\cref{eq:u_evol,eq:g_evol,eq:stress_s_relax}, i.e., $\mathbf{v}=\beta\mathbf{u}$, $\beta\nabla\cdot\mathbf{u}=\gamma$ and $\boldsymbol\sigma_s=\frac{\mu}{\beta}\Big(\nabla \mathbf{v}+\nabla\mathbf{v}^{\text{T}}-\frac{2}{d}(\nabla\cdot\mathbf{v})\mathbf{I}\Big)$. Together with Eq.~(\ref{eq:pressure_linear}) and $\nabla\cdot (\boldsymbol{\sigma}_s+\boldsymbol{\sigma}_p)=0$, we can derive the tissue behavior in mechanical equilibrium as viscous fluid 
\begin{equation}
\label{eq:steadyflow}
\frac{\mu}{\beta}\nabla^2\mathbf{v}=K\nabla (g-\frac{\gamma}{\beta}) -\frac{\mu}{\beta}(1-\frac{2}{d})\nabla\gamma,\quad \text{with the constraint}\quad \nabla\cdot{\mathbf{v}}=\gamma.
\end{equation}

When $d=2$, the last term on the right-hand side of Eq.~(\ref{eq:steadyflow}) vanishes, and our model is similar with the viscoelastic fluid model \cite{Streichan-2018-elife}, where the viscous tissue flow is driven by the force generated by the gradient of myosin-dependent constriction. However, in our linearized fluid system, the rate of force-generating activities $\gamma$ is the driver of the tissue dynamics, which results in a field of active volumetric or area increment $g$ and tissue flow $\mathbf{v}$. The tissue mechanics and flow can be modulated by the tissue fluidity $\beta$, together with the bulk modulus $K$ and effective viscosity $\mu/\beta$. The constraint $\nabla\cdot{\mathbf{v}}=\gamma$  implies incompressible tissue flow. However, the tissue can be under elastic compression or extension given that in general $J_e-1=\nabla \cdot \mathbf u -g+o(\epsilon)=\nabla \cdot \mathbf v/\beta -g+o(\epsilon)=\gamma/\beta -g+o(\epsilon)\neq 1$. 

\section{Growth of Compressible Tissues in Radial Symmetry}\label{sec:prescribed_growth}
Now we apply our model to consider the growth of compressible tissues in radial symmetry with tissue rearranging activities. In the following we will present the results for the growth of tissue spheroid ($d=3$). The results for the growth of tissue disc ($d=2$) are qualitatively similar to those for tissue spheroid and are included in Supplementary Materials. We will first find an analytic equilibrium solution of the linearized system. Then we will perform numerical simulations to the nonlinear system and compare the results with the linear solutions. 

\subsection{Equilibrium Solutions of the Linearized System}\label{subsec:Equil_soln}
We denote the solutions for the system of tissue spheroid as functions of the radial coordinate $r$ and time $t$, and write the system of Eqs. \cref{eq:u_evol,eq:g_evol,eq:pressure_linear,eq:stress_linear} in spherical coordinates $(r,\theta,\varphi)$ with the force balance Eq.~(\ref{eq:equil_eqn}), the boundary condition (\ref{eq:bc0}) at the tissue center $r=0$ and (\ref{eq:bc1}) at the free boundary $r=R(t)$, where $R(t)$ is the radius of tissue boundary at time $t$. Assuming that the active volumetric increment $g=J_g-1$, the adaptive displacement $u=r-y$, and $\partial u/\partial r$ are small, the linearized model of a growing tissue spheroid is given by

\begin{equation}\label{eq:linear_system}
\begin{dcases}
&\frac{\partial u}{\partial t}= v - \beta u, \\
&\frac{\partial g}{\partial t} = \gamma - \beta\Big(\frac{\partial u}{\partial r}+2\frac{u}{r}\Big), \\
&\frac{\partial \sigma_{rr}}{\partial r}+\frac{1}{r}(2\sigma_{rr}-\sigma_{\theta\theta}-\sigma_{\varphi\varphi})=0,\\
& \sigma_{rr}= \frac{4}{3}\mu \Big(\frac{\partial u}{\partial r} - \frac{u}{r}\Big) +K\Big(\frac{\partial u}{\partial r} + 2\frac{u}{r}-g\Big),\\
& \sigma_{\theta\theta}=\sigma_{\varphi\varphi}= \frac{2}{3}\mu \Big(\frac{u}{r}-\frac{\partial u}{\partial r}\Big) +K\Big(\frac{\partial u}{\partial r} +2 \frac{u}{r}-g\Big),
\end{dcases}
\end{equation}
subject to stress-free initial conditions, fixed boundary condition at the origin $r=0$, and free moving boundary condition at $r=R(t)$
\begin{equation}\label{eq:linear_bc}
\begin{dcases}
u(r,0)=0, \quad g(r,0) = 0,\\
v(0,t)=0,\\ 
\sigma_{rr}(R(t),t)=0,\quad 
\frac{dR}{dt} = v(R,t).
\end{dcases}
\end{equation}

We consider the case that there is an equilibrium solution when ${\partial u}/{\partial t}={\partial g}/{\partial t}={d R}/{dt}=0$ in the system \cref{eq:linear_system,eq:linear_bc}. We find the solution \begin{equation}\label{eq:equil_solution}
v(r)= \frac{1}{r^2}\int_0^{r} \gamma(\xi) \xi^2 d\xi,\quad g =\Big(1+\frac{4\mu}{3 K}\Big)\frac{\gamma(r) }{\beta}, \quad u(r)= \frac{v(r)}{\beta}.
\end{equation}
The first equation provides the existence criterion for an equilibrium tissue radius $R_{eq}$, which is the existence of a positive solution of    \begin{equation}\label{eq:R_eq}
G(R_{eq})=0, \quad \text{where}\quad G(r) := \int_0^{r} \gamma(\xi) \xi^2 d\xi.
\end{equation}
One can see $R_{eq}$ is solely determined by $\gamma$ through $G(r)$. Henceforth we call $G(r)$ the criterion function for the equilibrium size with respect to the growth rate $\gamma$. By assuming the continuity of $\gamma$, one can check there exists an equilibrium $R_{eq}$ only when $\gamma$ is partially positive and partially negative along the radial coordinate $r$.

We can also reconstruct the stresses and the elastic volume variation in equilibrium from the system \cref{eq:linear_system}: 
\begin{equation}\label{eq:stress_solution}
\sigma_{rr} = -\frac{4\mu}{\beta} \frac{G(r)}{r^3}, \quad
\sigma_{\theta\theta} =\frac{2\mu}{\beta}\Big(\frac{G(r)}{r^3}-\gamma(r)\Big),\quad
J_e = 1-\frac{4\mu}{3K}\frac{\gamma(r)}{\beta}.
\end{equation}

Thus, the stresses are regulated by the differential growth $\gamma(r)$ and modulated by the effective viscosity $\mu/\beta$ in equilibrium. It is noted that the stresses in this linearized system are independent on the bulk modulus $K$. The isotropic stress (pressure) $\sigma_p=(\sigma_{rr}+2\sigma_{\theta\theta})/3=-(4\mu/3\beta)\gamma(r)$ is also dissipated with large $\beta$, suggesting the relaxation of the deviatoric stress also relaxes the pressure.  This behavior is very different from the pressure relaxation dynamics in \cite{Ranft-2010-pnas} where the pressure relaxation can follow an independent rate close to homeostasis. We can compute the tissue density in equilibrium by $\rho=\rho_0/J_e\approxeq \rho_0(1+\frac{4\mu}{3K}\frac{\gamma(r)}{\beta})$. Thus, the spatial variation in density is modulated by $K/\mu$, the bulk-to-shear-modulus ratio, and $\beta/\gamma$, the fluidity-to-growth-rate ratio. Both large $K/\mu$ and large $\beta/\gamma$ result in small variation in density. The analytic equilibrium solutions of stress and density for the system of tissue disc ($d=2$) show similar qualitative behaviors (see Supplementary Materials). 

\subsection{Nonlinear Simulation with Prescribed Differential Growth Rate}\label{subsec:NL_simulation}
To investigate the regulation of tissue size and mechanics in the nonlinear regime, we perform numerical simulations to the nonlinear model Eqs.~\cref{eq:constitutive_law,eq:equil_eqn,eq:y_evol2,eq:Jg_evol2} with the fixed boundary condition (\ref{eq:bc0}) at $r=0$ and the free boundary condition (\ref{eq:bc1}) at $r=R(t)$

\begin{equation}\label{eq:nonlinear_sphere}
\begin{dcases}
&\frac{\partial y}{\partial t} + v\frac{\partial y}{\partial r}= \beta(r-y), \\ 
&\frac{\partial J_g}{\partial t} + v\frac{\partial J_g}{\partial r}= J_g\Big[\gamma + \beta\Big(3-(\frac{\partial y}{\partial r})^{-1}-2(\frac{y}{r})^{-1}\Big)\Big], \\ 
&\frac{\partial \sigma_{rr}}{\partial r}+\frac{1}{r}(2\sigma_{rr}-\sigma_{\theta\theta}-\sigma_{\varphi\varphi})=0, \\
&\sigma_{rr}= \frac{2}{3}\mu J_g\Big((\frac{\partial y}{\partial r})^{-\frac{1}{3}}(\frac{y}{r})^{\frac{10}{3}}-(\frac{\partial y}{\partial r})^{\frac{5}{3}}(\frac{y}{r})^{\frac{4}{3}}\Big)+K\Big(J_g^{-1}(\frac{\partial y}{\partial r})^{-1}(\frac{y}{r})^{-2}-1\Big), \\ 
&\sigma_{\theta\theta}= \frac{1}{3}\mu J_g\Big((\frac{\partial y}{\partial r})^{\frac{5}{3}}(\frac{y}{r})^{\frac{4}{3}}-(\frac{\partial y}{\partial r})^{-\frac{1}{3}}(\frac{y}{r})^{\frac{10}{3}}\Big)+K\Big(J_g^{-1}(\frac{\partial y}{\partial r})^{-1}(\frac{y}{r})^{-2}-1\Big),\\
& \sigma_{\varphi\varphi}= \sigma_{\theta\theta},
\end{dcases}
\end{equation} 
subject to the stress-free initial condition, the fixed boundary condition at $r=0$, and free moving boundary condition at $r=R(t)$
\begin{equation} 
\begin{dcases}
&y(r,0)=r, \quad J_g(r,0) = 1, \\
&v(0,t)=0,\\ 
&\sigma_{rr}(R,t)=0,\quad
\frac{dR}{dt} = v(R,t).
\end{dcases}
\end{equation}
To solve this nonlinear system, we first introduce the change of variable $r'=r/R(t)\in[0,1]$ to the original system. The change of variable transforms the system from a moving domain $[0,R(t)]$ to a fixed domain $[0,1]$. We discretize the new system by finite difference in space and time. The discretized system at each time point is solved by a semi-implicit iterative method. The details of the numerical method is not the core of this paper and will be presented in a different manuscript.

\begin{figure}[!t]
	\centering
	\includegraphics[width=1\textwidth]{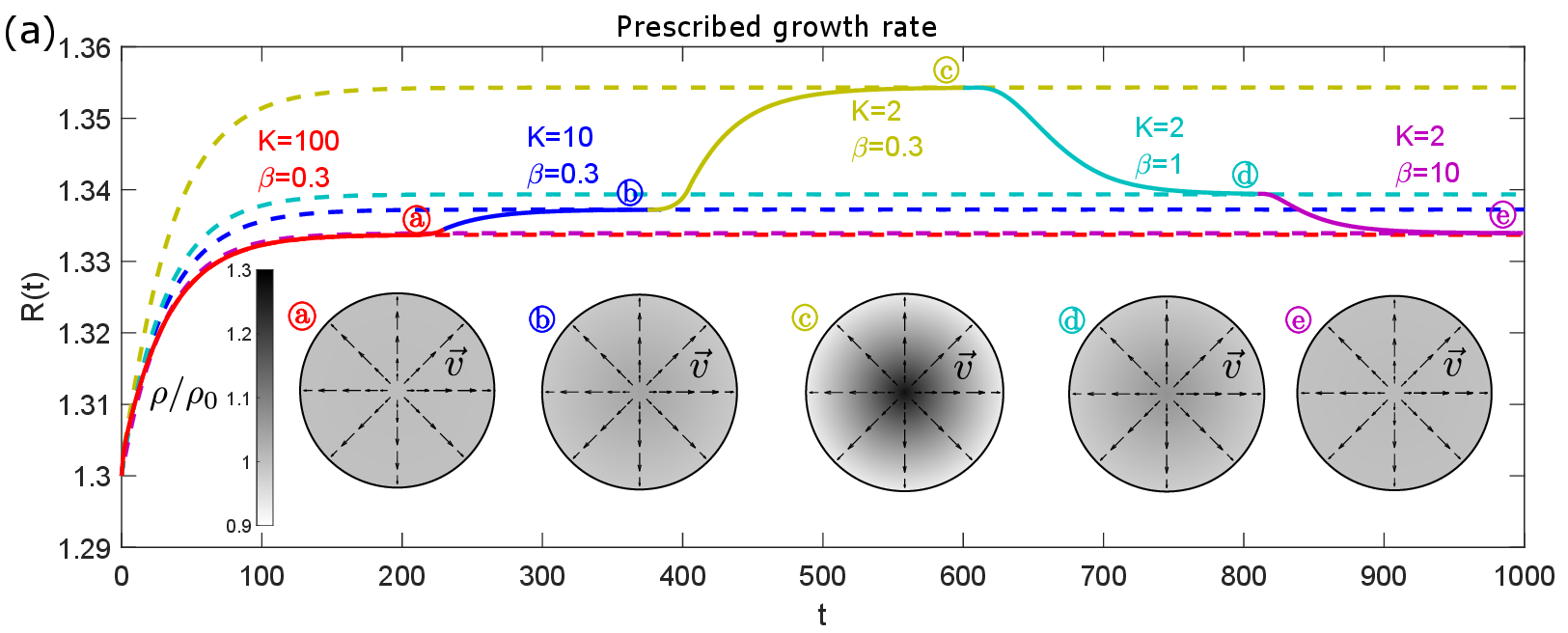}\\
	\hspace{0.05\textwidth}
	\includegraphics[width=0.35\textwidth]{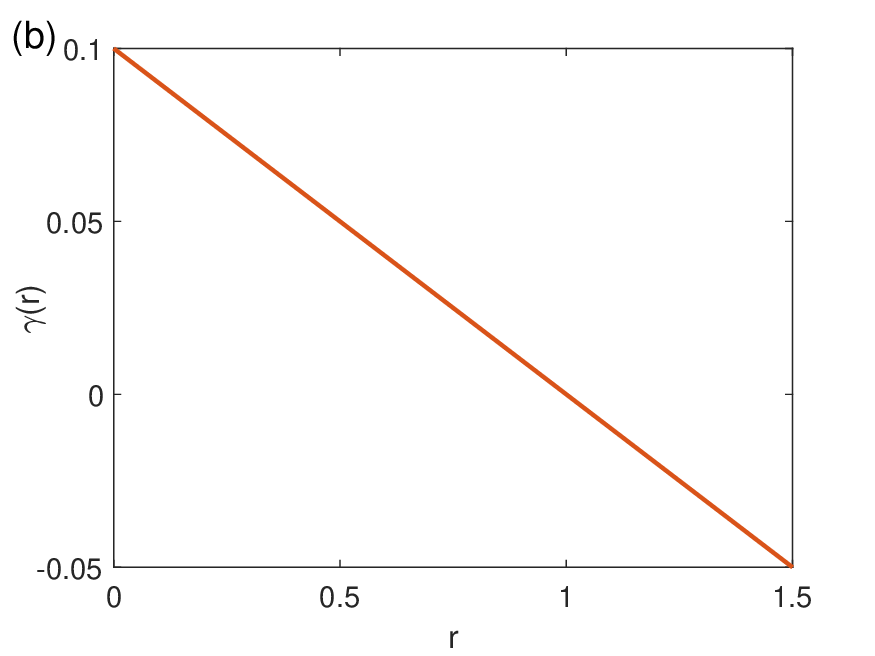}
	\hspace{0.05\textwidth}
	\includegraphics[width=0.35\textwidth]{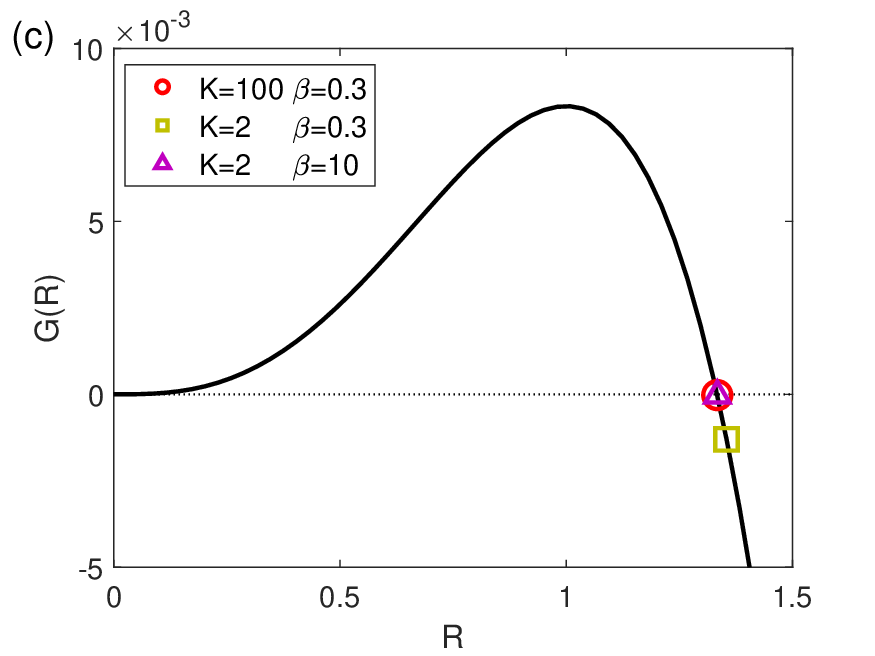}	
	\caption{(a) Evolution of tissue radius $R(t)$ with dynamical modulation of $\beta$ and $K$ (solid line) and with different sets of constant $K$ and $\beta$ (dashed lines), with prescribed $\gamma(r)=0.1(1-r)$. The gray-scale color map in the disc panels (a-e) shows the scaled density $\rho/\rho_0$ profile within tissue spheroid in equilibria for each set of parameters $\beta$ and $K$. The arrows indicate the direction and relative magnitude of the tissue velocity. (b) The spatial profile of prescribed $\gamma(r)$. The location $r=0$ corresponds to the tissue center. (c) Criterion function $G(R)$ with respect to $\gamma(r)=0.1(1-r)$. The equilibrium size $R_{eq}$ is marked (see legend) for different values of $K$ and $\beta$. The equilibrium size $R_{eq}$ deviates from the positive root of $G(R)$ for highly compressible and weakly fluidized ($K=2$ and $\beta=0.3$) tissues. }\label{fig:regulation}
\end{figure}

From now on, we set $\mu=1$ without loss of generality, and then $K$ stands for the relative tissue compressibility $K/\mu$. We consider a prescribed differential growth rate $\gamma(r)=0.1(1-r)$ (Fig.~\ref{fig:regulation}(b)), which indicates a differential growth with cell proliferation near the tissue center and apoptosis near the boundary. This choice of $\gamma$ gives the equilibrium tissue size of $R_{eq}=4/3$ for the linear system from Eq.~(\ref{eq:R_eq}). We simulate the growth of tissue from a smaller size $R=1.3$ until it reaches its mechanical equilibrium state. 

\subsubsection{Nonlinear Effect on Equilibrium Tissue Size}\label{subsec:tissue_size}
The evolution of tissue size $R(t)$ with different values of $K$ and $\beta$ is shown in Fig.~\ref{fig:regulation}(a) (dashed lines). The results show that the equilibrium size of tissue for the nonlinear system, depending on $K$ and $\beta$, can deviate from the linear solution $R_{eq}=4/3$. When $K$ or $\beta$ is large, the equilibrium size of nonlinear system is close to the linear solution; however, when $K$ and $\beta$ are both small, the tissue reaches a different equilibrium size. The disc panels in Fig.~\ref{fig:regulation}(a) show the spatial profile of the rescaled tissue density $\rho/\rho_0$ (by gray-scale color maps) and tissue flow $v$ (by arrows) in equilibria. Due to the negative gradient of $\gamma$, the tissue density is larger near the center and smaller near the boundary, and tissue flows from the center to the boundary. It shows that the variation in density becomes significant when both $K$ and $\beta$ are small, in a good agreement with the prediction from linear-model solution in equilibrium. 
Moreover, if we update $K$ or $\beta$ after the tissue reaches an equilibrium size, $R(t)$ will evolve to the new equilibrium corresponding to the updated values of $K$ and $\beta$, as shown in Fig.~\ref{fig:regulation}(a) (solid line). We have further verified that these results are independent of the initial size choice $R(0)$. Thus, we demonstrate that given $\gamma(r)$, the equilibrium tissue radius $R_{eq}$ also depends on $K$ and $\beta$. The nonlinear simulation results for tissue disc ($d=2$) display similar model behavior (see Supplementary Materials for more details).

We can understand why $K$ and $\beta$ modulate $R_{eq}$ in the nonlinear model by the following analysis. When tissue is in mechanical equilibrium, its elastic volume variation $J_e=J/J_g$, or equivalently, its density $\rho=\rho_0/J_e$ does not change locally (with respect to Eulerian coordinates), i.e., $\partial \rho/\partial t= \rho(\gamma-\nabla\cdot \mathbf{v}) - \mathbf{v} \cdot \nabla \rho =0$. By rewriting this relation into $\gamma =\nabla\cdot \mathbf{v} +(1/\rho) \mathbf{v}\cdot \nabla \rho$ and integrating it in a tissue spheroid  ($d=3$) or disc ($d=2$), we obtain \begin{equation}\label{eq:R_eq_NL}
G(R_{eq}) =  \int_0^{R_{eq}} \frac{1}{\rho} \Big(v\frac{\partial \rho}{\partial r}\Big) r^{d-1} dr, \quad \text{with } G(R)=\int_0^{R} \gamma r^{d-1} dr,
\end{equation}
where we have used the divergence theorem and  $v(0,t)=0$ and $v(R_{eq},t)=0$.

For strictly incompressible case, where $\rho\equiv\rho_0$,  Eq.~(\ref{eq:R_eq_NL}) is reduced to $G(R_{eq})=0$, where $R_{eq}$ is solely determined by $\gamma$ via the criterion function $G$, as in the linear case by Eq.~(\ref{eq:R_eq}). This is also the case when the tissue is nearly incompressible or highly fluidized (e.g., $K=100$ or $\beta=10$), $\rho$ is (almost) uniformly constant such that $\partial \rho/\partial r\approxeq 0$. However, when the tissue is highly compressible and weakly fluidized (e.g., $K=2$ and $\beta=0.3$), the right hand side of Eq.~(\ref{eq:R_eq_NL}) becomes negative since the nonlinear advection term $v{\partial \rho}/{\partial r}<0$ everywhere except at $r=0$ and $r=R_{eq}$. Thus the solution of $R_{eq}$ to Eq.~(\ref{eq:R_eq_NL}) is located at a different value where $G<0$ (see Fig.~\ref{fig:regulation}(c)). In \cite{Wu-2019-stress} where the strictly incompressible tissue is considered, the tissue size can be modulated by external physical confinement through mechanical feedback mechanisms. Here, we show that the tissue material properties can modulate the tissue size without considering external compression or mechanical feedback mechanisms. To the best of our knowledge, this size-regulating behavior of tissues in a mechanically free growing environment due to its internal mechanical property and rearranging activity has not been predicted by previous models.

\begin{figure}[!b]
	\centering
	\textbf{\scriptsize Prescribed growth factor}\\
	\includegraphics[width=0.32\textwidth]{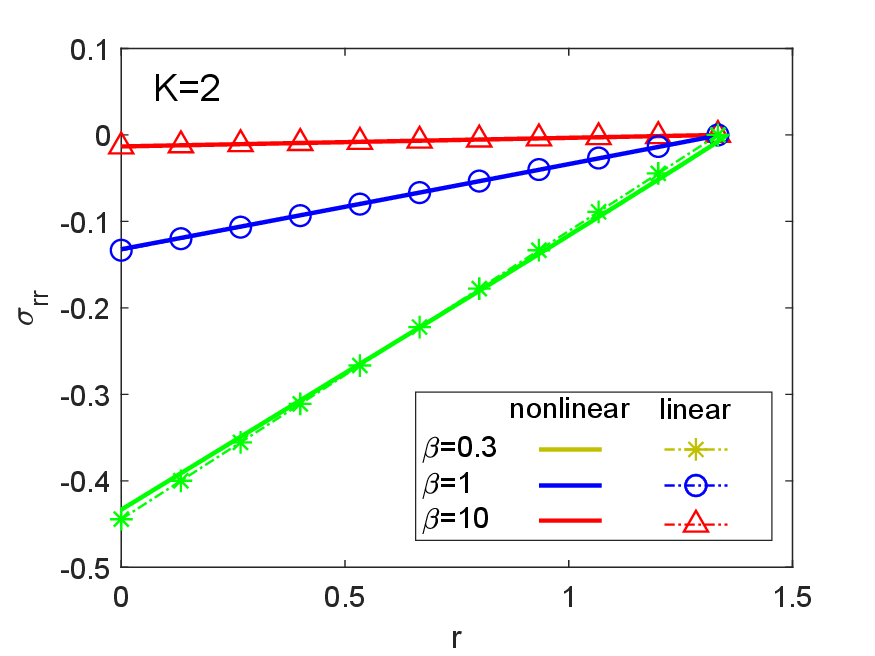}
	\includegraphics[width=0.32\textwidth]{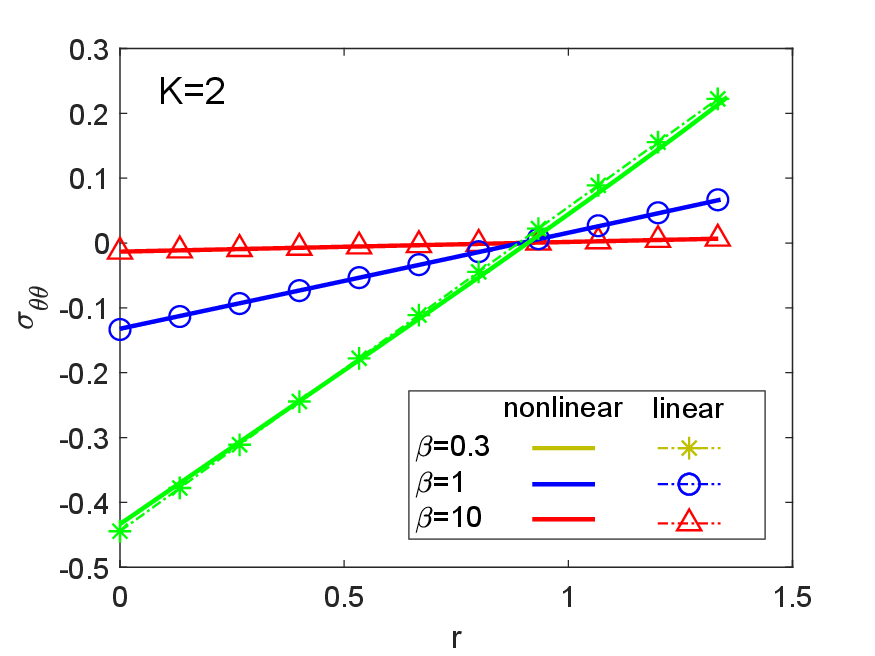}
	\includegraphics[width=0.32\textwidth]{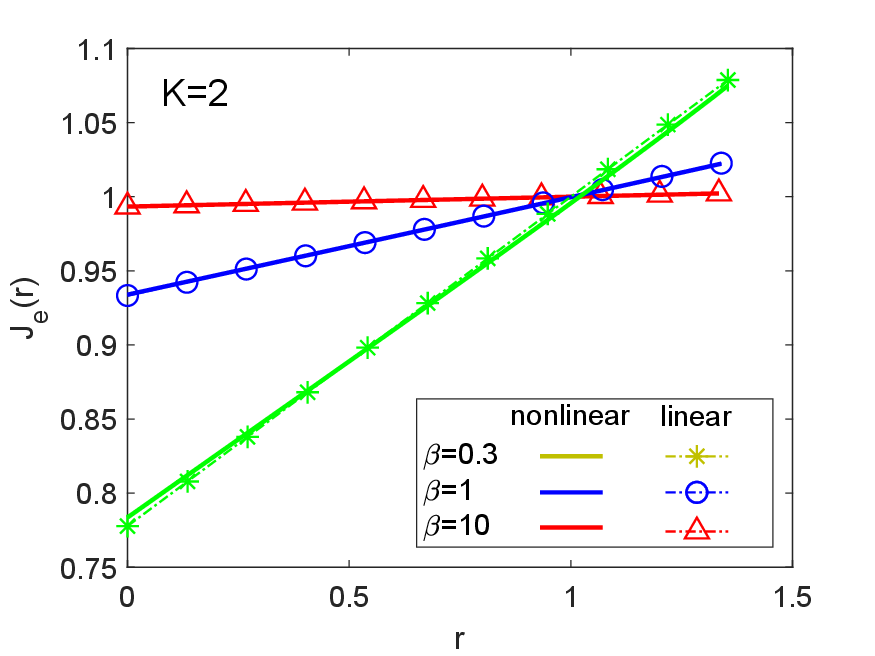}
	\caption{The Cauchy stress ($\sigma_{rr}$ and $\sigma_{\theta\theta}$) and elastic volume variation $J_e=\rho_0/\rho$ in equilibrium for different $\beta$ with prescribed $\gamma(r)=0.1(1-r)$. Solid lines represent nonlinear simulation results; markers represent linear analytic solutions. }\label{fig:beta_regulation}
\end{figure}

\subsubsection{Regulation of Tissue Mechanics}\label{subsec:regulation}
Given the prescribed $\gamma(r)=0.1(1-r)$, the spatial distributions of the Cauchy stress components $\sigma_{rr}$, $\sigma_{\theta\theta}$, and the local elastic volume variation $J_e= (\rho/\rho_0)^{-1}$ are almost linear, under different combinations of $\beta$ and $K$. Their distributions for $K=2$ and $\beta =10,1,0.3$ are shown in Fig.~\ref{fig:beta_regulation} (solid lines). Since the nonlinear solutions of these variables are qausi-linear in $r$, they matches closely with the linear model solutions (\ref{eq:stress_solution}) with different values of $K$ and $\beta$ (dashed lines in Fig.~\ref{fig:beta_regulation}). Overall, the tissue is under compression near the center (where the density is larger) and under extension near the boundary (where the density is smaller). When the tissue fluidity is large (large $\beta$), the Cauchy stress components $\sigma_{rr}$ and $\sigma_{\theta\theta}$ are dissipated and the elastic volume variation $J_e$ is (almost) uniformly $1$. When the tissue fluidity is small (small $\beta$), the gradients of $\sigma_{rr}$, $\sigma_{\theta\theta}$, and   $J_e$  are more significant. We also find that the stress in equilibrium is independent of $K$, which is consistent with
the analytical linear solution (\ref{eq:stress_solution}). Similar behavior in tissue disc ($d=2$) is shown in Supplementary Materials. 

\section{Application to Tissue Growth Coupled with Chemical Signals}\label{sec:coupling_growth}
Now we apply our nonlinear model to study the growth of radially symmetric tissue coupled with chemical signals. We will show the results for tumor spheroid growth ($d=3$) in the following, and the results for epithelial disc growth ($d=2$) are included in Supplementary Materials. 
We consider that the growth rate $\gamma$ is no longer prescribed as an {\it a priori} function but is coupled with a field of growth factor or nutrient that follows the reaction-diffusion process in the growing tissue domain. Supposing that $c(r,t)$ is the concentration of growth factor in the tumor spheroid, its dynamics follows
\begin{equation}\label{eq:diffusion}
\begin{dcases}
&\frac{1}{\lambda}\frac{\partial c}{\partial t} = L^2\frac{1}{r^2}\frac{\partial}{\partial r}\Big(r^2\frac{\partial c}{\partial r}\Big)- c,  \quad  \text{for $0< r< R(t)$,}\\
&c(R(t),t)=1, \quad \frac{\partial c}{\partial r}(0,t)=0,
\end{dcases}
\end{equation}
where we assume Dirichlet boundary condition at the boundary and no flux boundary condition at the center such that the growth factor diffuses from the boundary to the central region. The turnover rate of the growth factor $\lambda$ determines the characteristic time of the reaction-diffusion process and $L=\sqrt{D/\lambda}$ is the diffusion length, where $D$ is diffusion coefficient. Assuming $\lambda\gg \gamma$, we can solve $c$ by the quasi-steady-state approximation of Eq.~(\ref{eq:diffusion}) for tissue radius $R(t)$ at each time point. The obtained solution $c_R(r)$ is time-dependent via the moving domain boundary $R(t)$:
\begin{equation}
c_R(r):=c(r,t) = \frac{R(t)}{\sinh(R(t)/L)}\frac{\sinh(r/L)}{r}.
\end{equation}

\begin{figure}[!h]
	\centering
	\includegraphics[width=0.9\textwidth]{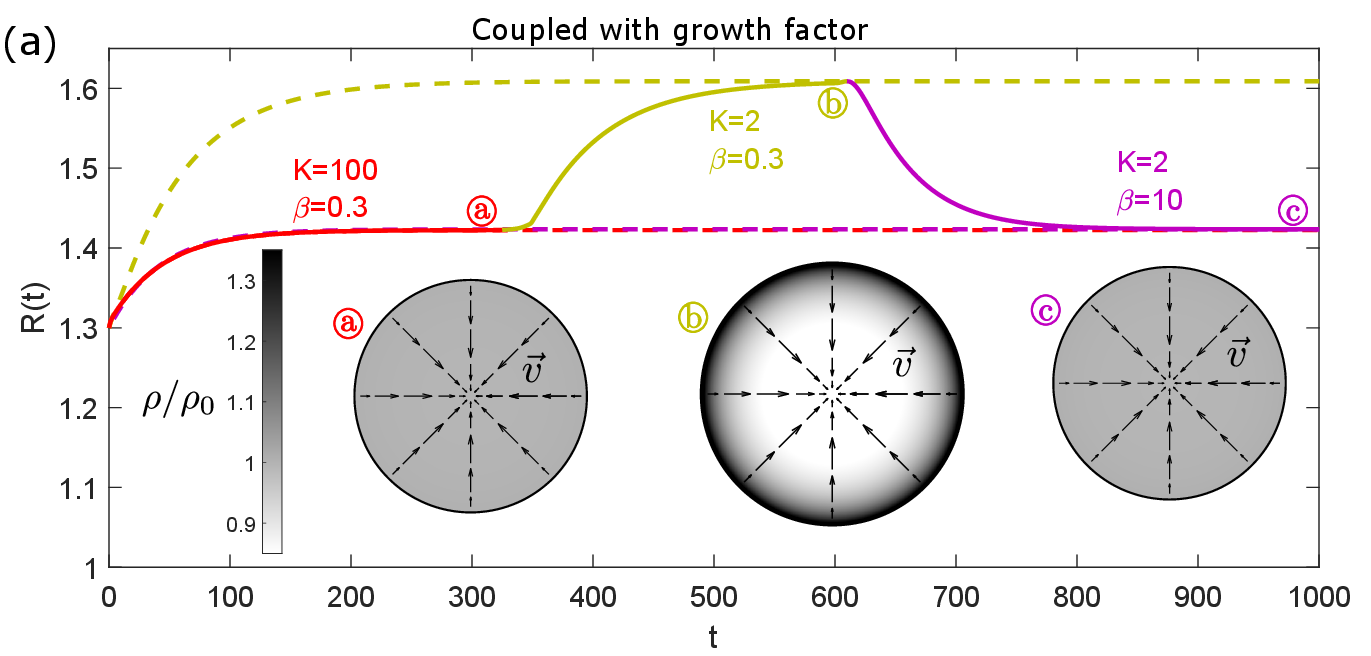}\\
	\hspace{0.05\textwidth}
	\includegraphics[width=0.4\textwidth]{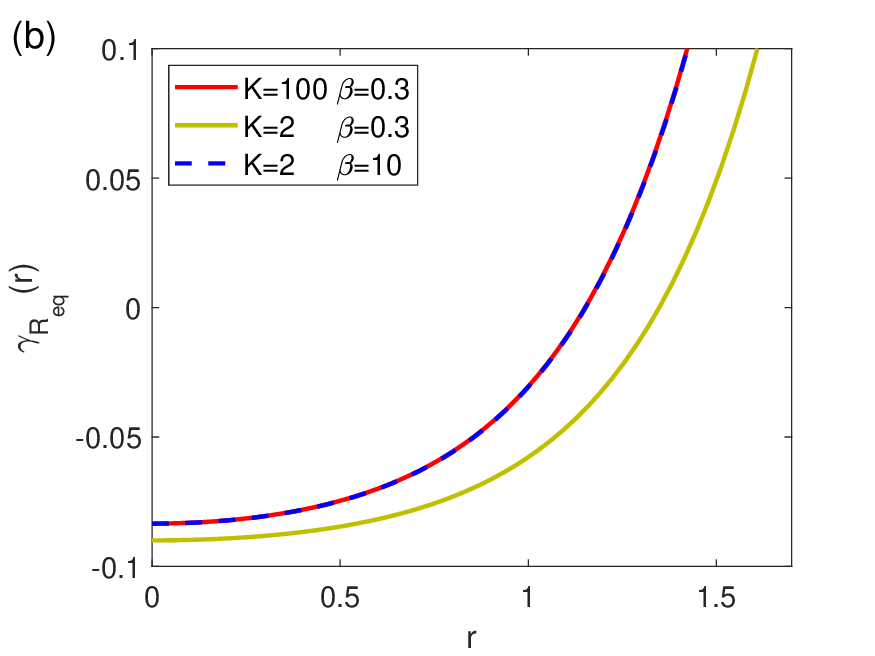}
	\hspace{0.05\textwidth}
	\includegraphics[width=0.4\textwidth]{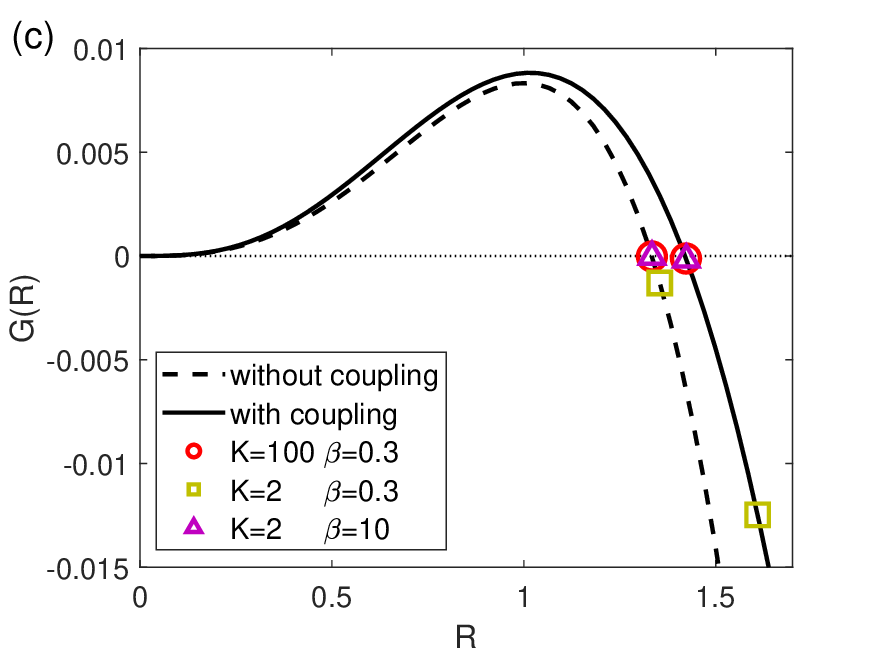}	
	\caption{(a) Evolution of tumor size $R(t)$ with dynamical modulation of $\beta$ and $K$ (solid line) and with different sets of constant $K$ and $\beta$ (dashed lines), when $\gamma$ is coupled with growth factor. Inset: the colored disk shows the scaled density $\rho/\rho_0$ profile within tumor spheroid in equilibria (a-c) for each set of parameters $\beta$ and $K$. The arrows indicate the direction and dimensionless magnitude of the velocity of tissue flow. (b) The spatial profile of $\gamma_{R_{eq}}(r)$ in equilibria when coupled with growth factor for different $K$ and $\beta$. (c) Criterion function $G(R)$ for tumor equilibrium size when $\gamma$ is prescribed {\it a priori} (solid line) and coupled with growth factor (dashed line). The equilibrium size $R_{eq}$ is marked (see inset) for different values of $K$ and $\beta$. The equilibrium size $R_{eq}$ deviates from the root for highly compressible and weakly fluidized ($K=2$ and $\beta=0.3$) tumors. This deviation is pronounced when $\gamma_R$ is coupled with growth factors. }\label{fig:coupling}
\end{figure}

We consider that the local rate of active growth is coupled with the local concentration $c_R(r)$ in the simplified form of 
\begin{equation}
\label{eq:gamma_R}
\gamma_R(r) =  c_R(r) \lambda_{p}- \lambda_{a},
\end{equation}
where $c_R\lambda_{p}$ and $\lambda_{a}$ are the local rates of cell proliferation and apoptosis, respectively. Typical $\gamma_R(r)$ decreases from the boundary to the center of the tissue, and for a fixed location $r_0$, $\gamma_R(r_0)$ further decreases as $R(t)$ increases. Figure~\ref{fig:coupling}(b) shows $\gamma_{R_{eq}}(r)$ in equilibria with different $R_{eq}$. We perform simulations for the growth of tumor spheroid with $\gamma=\gamma_R(r)$, with the parameter values $L=0.3$, $\lambda_p=0.2$ and $\lambda_a=0.1$. 

\begin{figure}[!h]
	\centering
	\textbf{\scriptsize Coupled with growth factor}\\
	\includegraphics[width=0.5\textwidth]{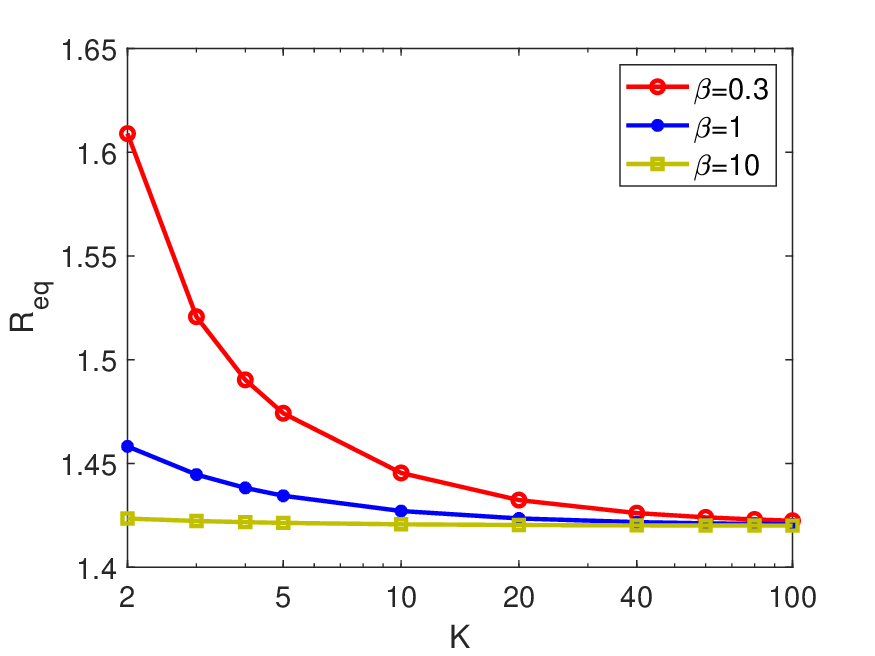}
	\caption{The equilibrium tumor size $R_{eq}$ for different $\beta$ and $K$.}\label{fig:R_K}
\end{figure}

The evolution of tumor spheroid with $\gamma_R$ for different $K$ and $\beta$ is shown in Fig.~\ref{fig:coupling}(a). The equilibrium tissue size can be modulated by $K$ and $\beta$ due to the same mechanism as explained in Sec. \ref{subsec:tissue_size}. We show the equilibrium tumor size $R_{eq}$ for different $\beta$ and $K$ in Fig.~\ref{fig:R_K}. When the tumor is nearly incompressible ($K=100$) or highly fluidized ($\beta=10$), $R_{eq}$ coincides with the root of the criterion function $G(R)$ (Fig.~\ref{fig:coupling}(c)); when the tumor is highly compressible (small $K$) and weakly fluidized (small $\beta$), $R_{eq}$ deviates from the root of $G(R)$ and increases to the location where $G(R)<0$. Compared to the case of prescribed $\gamma$ (Fig.~\ref{fig:regulation}(a)), the variation in $R_{eq}$ for different $K$ and $\beta$ becomes significant in the case where  $\gamma=\gamma_R$ is coupled with $c_R(r)$. 

\begin{figure}[!h]
	\centering
	\textbf{\scriptsize Coupled with growth factor}\\
	\includegraphics[width=0.32\textwidth]{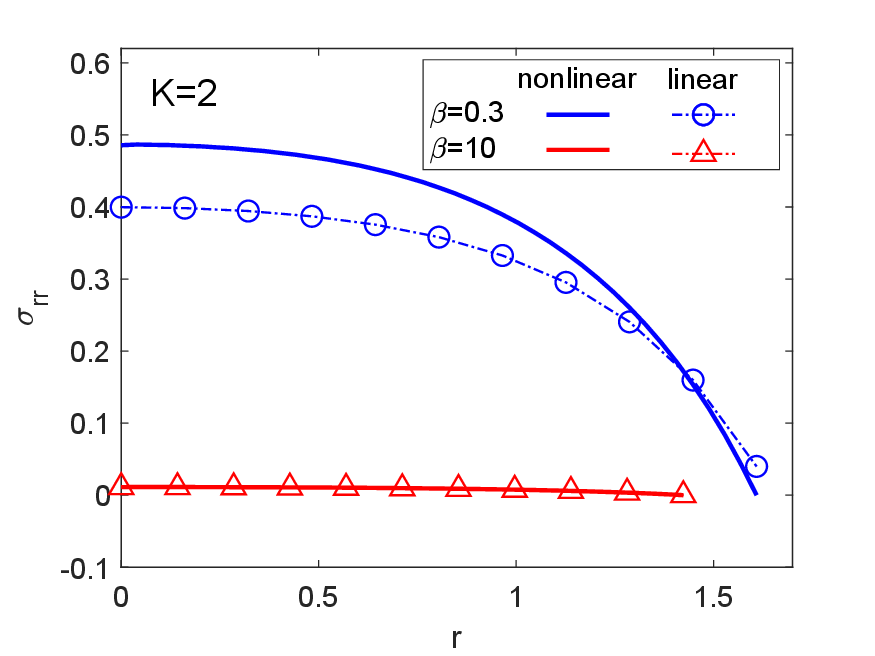}
	\includegraphics[width=0.32\textwidth]{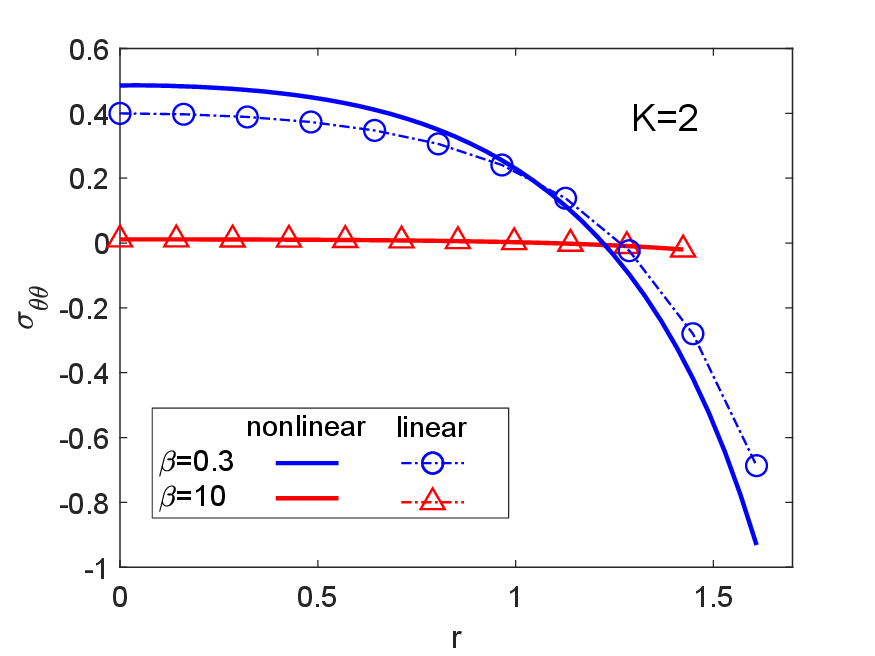}
	\includegraphics[width=0.32\textwidth]{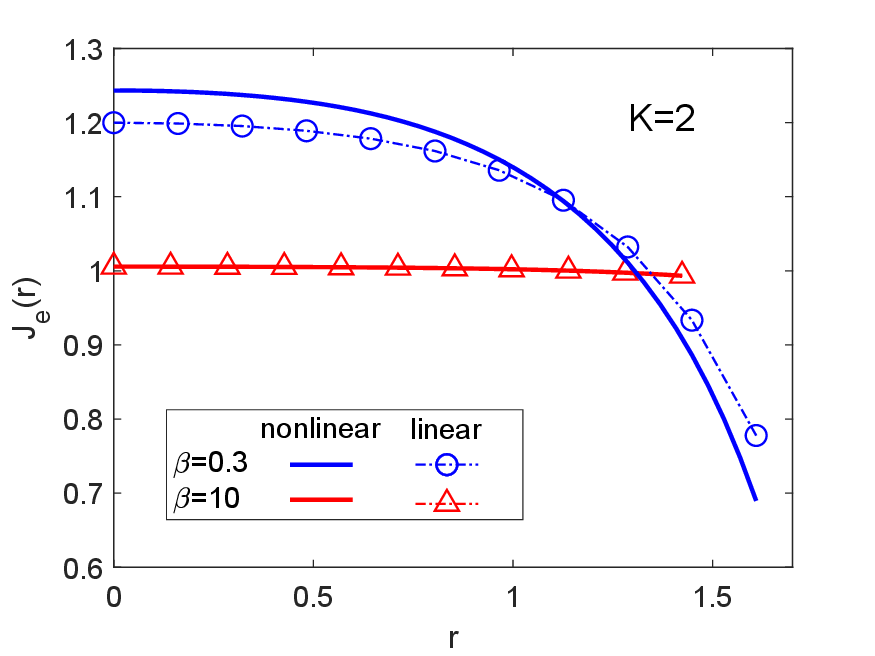}	
	\caption{The Cauchy stress ($\sigma_{rr}$ and $\sigma_{\theta\theta}$) and elastic volumetric change $J_e$ in equilibrium for different $\beta$ when $\gamma$ is coupled with growth factor. Solid lines represent nonlinear simulation results; markers represent linear analytic solutions. }\label{fig:beta_regulation2}
\end{figure}

The positive gradient of $\gamma_R$ results in a positive gradient in tissue density in equilibrium and induces inward tissue flow from the boundary to the center (see disc panels in Fig.~\ref{fig:coupling}). Therefore, the tumor tissue is under tension near the center and compression near the boundary (Fig.~\ref{fig:beta_regulation2}). The spatial distributions of the Cauchy stress components $\sigma_{rr}$, $\sigma_{\theta\theta}$, and the local elastic volume variation $J_e= (\rho/\rho_0)^{-1}$ with larger fluidity  (e.g., $\beta=10$) match well with the prediction from the linear solution Eq. (\ref{eq:stress_solution}). For smaller fluidity (e.g., $\beta=0.3$), the nonlinear solutions obviously deviate from the linear solutions due to the stronger nonlinear effect.  Moreover, we find that the stresses, unlike in the case of prescribed $\gamma$, can be also modulated by $K$, which is not the case for the linear solutions; see Supplemental Materials for more details illustrated with $d=2$.

\section{Summary and Discussion}\label{sec:discussion}
In this work, we presented a nonlinear biomechanical model for growing compressible tissues which couples the finite elasticity and growth with spatially-dependent growth-promoting factors in an Eulerian frame. We have used the reference map techniques \cite{Cottet-2008-referencemap,Kamrin-2012-referencemap}, which track the inverse of the motion. This formulation is simpler than other Eulerian formulations based on tracking the evolution equations of tensor variables. For example, previous Eulerian-frame works described the evolution of deformation gradient for viscoelastic fluids \cite{Liu2001ARMA,Lin2005cpam} and the elastic deformation gradient for deformable solids considering surface growth \cite{Naghibzadeh2021jmps}.

Following \cite{Wu-2019-stress}, we have introduced a rearrangement rate $\beta$ (called tissue fluidity) to the reference map equation to account for tissue rearranging activities in compressible tissue. In \cite{Wu-2019-stress}, we have considered the tissue volume reductions as active processes of water efflux \cite{guo2017cell}. Here, we consider this tissue compressibility by its bulk modulus $K$, focusing on the solid structure that provides the elastic property approximated by the compressible neo-Hookean constitutive law. Without the constraint of strict incompressibility, we are able to demonstrate that the rearranging activity indeed dissipates the elastic energy in radial symmetry (where rotation can be neglected).

When the rearranging activities, such as cell topological transition and cytoskeletal reorganization, are much faster than the growth  ($\gamma/\beta\ll 1$), we can approximate the nonlinear model by its linearized version. In the linear model, we have found Maxwell-type relaxation for the stress dynamics, where the relaxation time is scaled with $1/\beta$. This linear model is similar to the Maxwell-type viscoelastic model in \cite{Ranft-2010-pnas}, yet their relaxation time is growth-dependent (through anisotropic growth). If we consider the characteristic time $T\sim 1/\gamma\gg1/\beta$, our model is reduced to an active viscous fluid model, which is similar to previous viscous fluid models \cite{bosveld2012mechanical,Streichan-2018-elife}.  In our model, we have found that the effective tissue viscosity is given by $\mu/\beta$, regulated by the shear modulus and the rearrangement rate. This relation connects the effective viscosity of tissues directly to the rearranging activities and elastic modulus when the characteristic time  $T\sim 1/\gamma\gg1/\beta$. 

\begin{figure}[!h]
	\centering
		\includegraphics[width=0.33\textwidth]{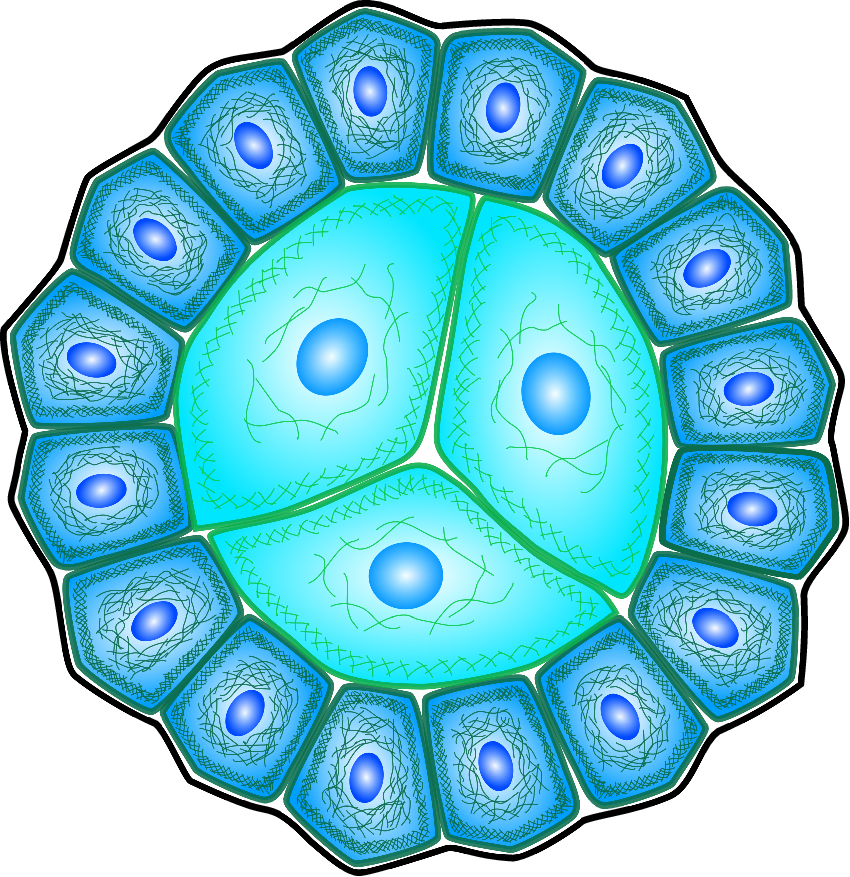}
			\includegraphics[width=0.30\textwidth]{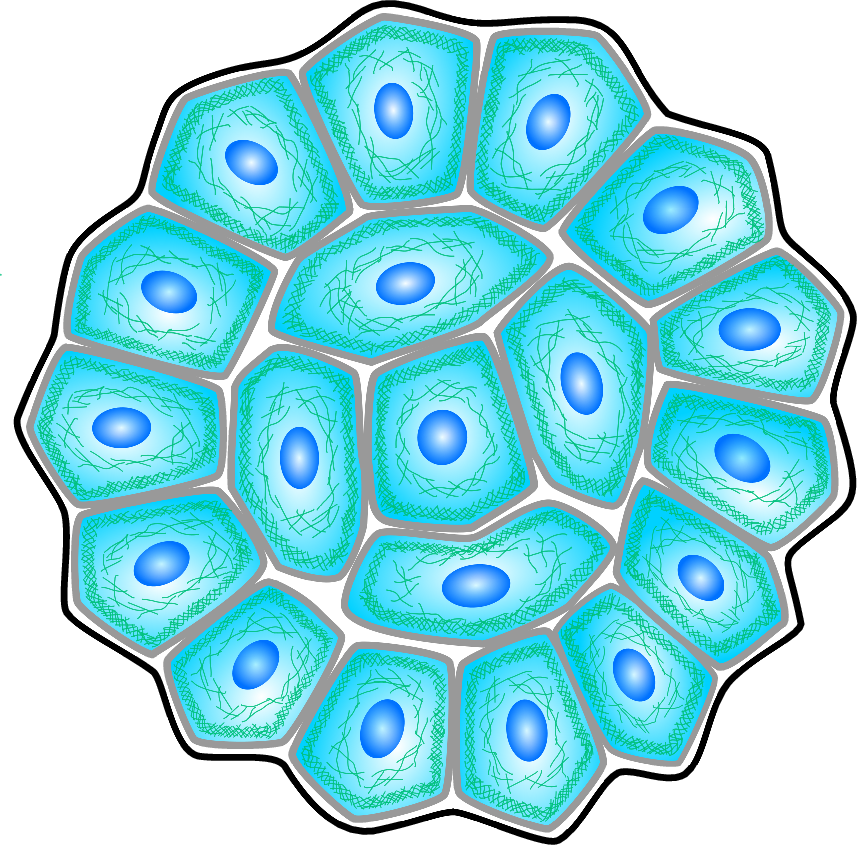}	
	\includegraphics[width=0.28\textwidth]{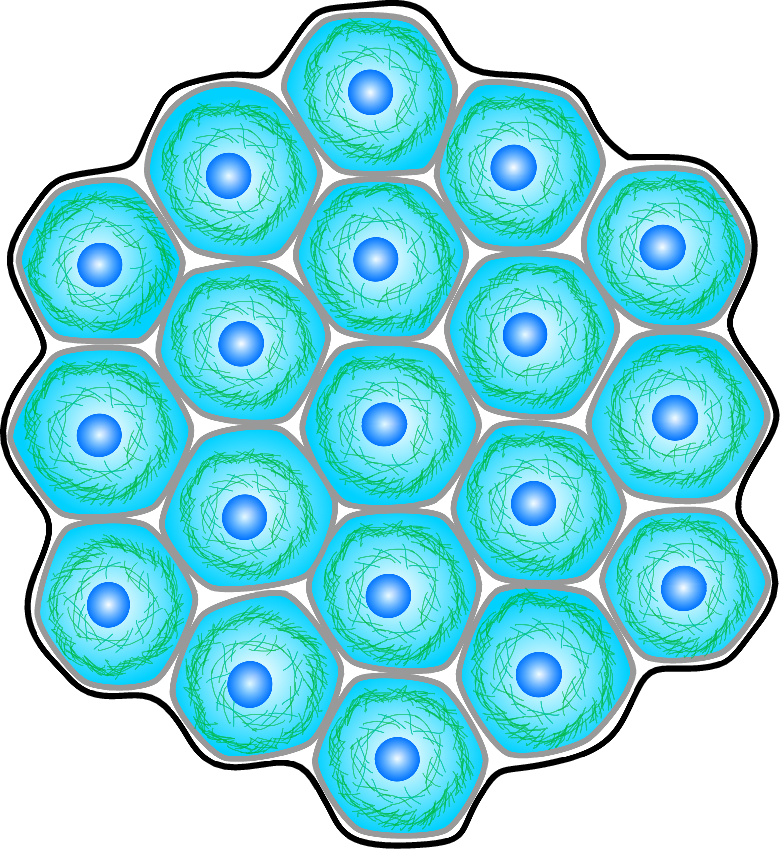}	
	\caption{For weakly fluidized compressible tissue (left), the equilibrium size is larger than the incompressible (middle) or highly fluidized (right) tissue size, due cell size change in response to the internal stress. See text for more details.}
	\label{scheme}
\end{figure}

This reference-map based viscoelastic compressible tissue growth model is efficient to investigate in radially symmetric growth (e.g., tumor spheroids and epithelial discs growth), thanks to its relative simple formulation. We have applied this model to study the growth of compressible tissue spheroid and disc. When the ratio between tissue fluidity and growth rate $\beta/\gamma$ is large, we have presented the analytical steady-state solution of the linear model, and we have found that the tissue size is regulated solely by the spatial distribution of growth rate $\gamma(r)$ in radially symmetric tissues and the Cauchy stress distribution is regulated by the viscosity $\mu/\beta$ and $\gamma(r)$. In the nonlinear regime, we have found both the compressibility $K/\mu$ and the tissue fluidity $\beta$ can influence the tissue equilibrium size in the mechanically free growing environment. When the tissue fluidity is small, as the tissue becomes more compressible, the tissue equilibrium size becomes larger. We have explained this effect as a nonlinear effect from  the advection of the density gradient $v\frac{\partial \rho}{\partial r}$. Intuitively, we can also understand this result by the following. For weakly fluidized compressible tissue with larger proliferation at the boundary, we have higher density at the boundary and lower density at the center (e.g., see Fig.~\ref{fig:coupling}). If we assume the cell density (cell number per volume) is proportional to the mass density (mass per volume), this means there are more proliferating cells at the tissue boundary which expand their sizes while flowing to the tissue center (see Fig.~\ref{scheme}). However, if the tissue is highly incompressible, cells will only change their shape but not their size in the tissue flow. With high fluidity, the stress level is low, thus cells do not change their size or shape during the flow. Thus, for highly incompressible or fluidized tissues, the equilibrium size will not be affected by the mechanics. This behavior is not observed in the incompressible model \cite{Wu-2019-stress}. However, if an increased volume growth induced by tensile stress has been considered, the incompressible model may generate similar results.

This Eulerian-frame model is ready to incorporate the coupling between the growth rate and stresses to investigate how tissue mechanical properties (e.g., compressibility and fluidity) influence tissue size regulation in radially symmetric growth (e.g., tumor spheroids and epithelial discs growth) via mechanical feedback mechanisms \cite{Shraiman-2005-pnas,Wu-2019-stress}.
Moreover, the presented model has the potential to describe the coupling between tissue growth and mechanics in higher dimensions beyond radial symmetry. In future, we expect to extend the current Eulerian model to consider rearranging activities on higher dimensions beyond radial symmetry. For example, it is interesting to study the effect of tissue fluidity on symmetry-breaking events to explain sophisticated shape formations \cite{Li-2011-jmps,shyer2013villification,amar2013anisotropic}.

\section*{Acknowledgment}
M.W acknowledges funding from NSF-DMS-2012330.

\bibliographystyle{siamplain}
\bibliography{myreference}

\end{document}


\title{Supplemental Material for \\An Eulerian nonlinear elastic model for compressible and fluidic tissue with radially symmetric growth} 

\author{Chaozhen Wei}
\email{cwei4@wpi.edu}
\affiliation{Department of Mathematical Sciences, Worcester Polytechnic Institute, Worcester, MA 10605 USA
}
\author{Min Wu}
\email[To whom correspondence should be addressed.\\E-mail: ]{englier@gmail.com}
\affiliation{Department of Mathematical Sciences, Worcester Polytechnic Institute, Worcester, MA 10605 USA
}

\date{\today}

\maketitle

\section{Dynamics with Tissue Rearrangement}\label{sec:tissue-rearrange}
\begin{figure}[!h]
	\centering
	\includegraphics[width=0.8\textwidth]{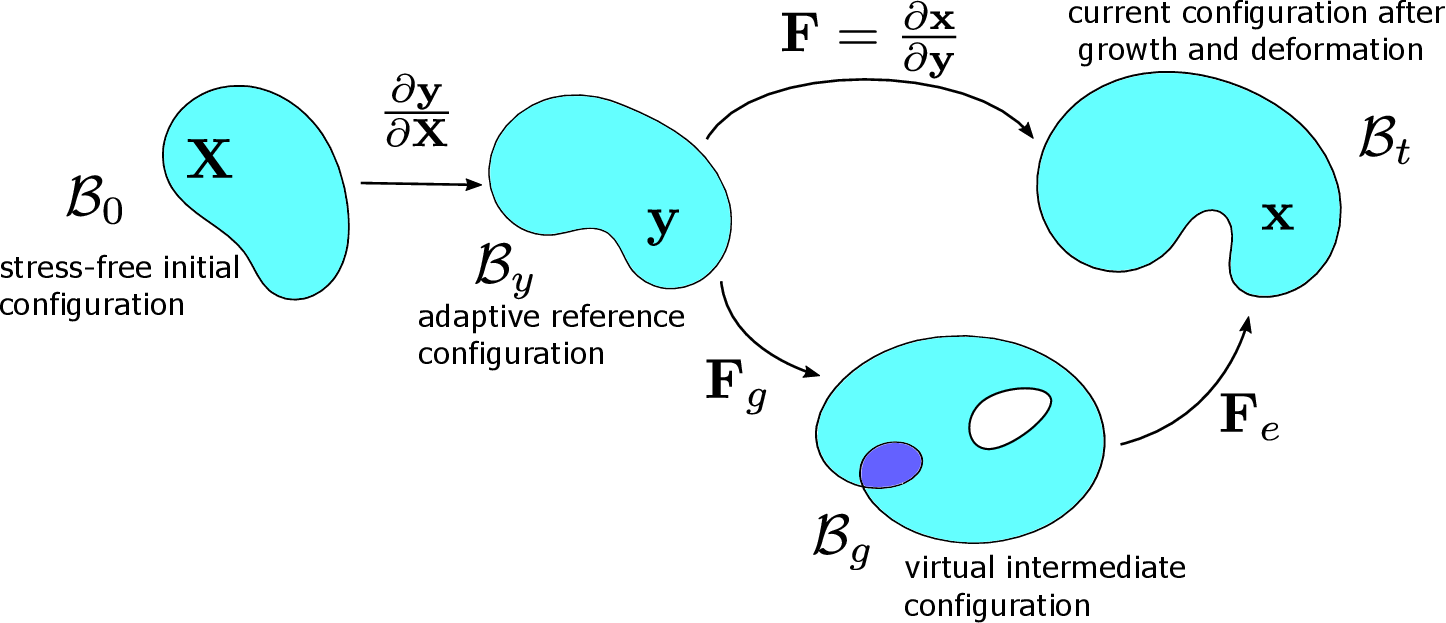}
	\caption{Decomposition of the geometric deformation gradient into growth stretch tensor and elastic deformation tensor. Loop $\mathcal{B}_0\rightarrow\mathcal{B}_y\rightarrow\mathcal{B}_g\rightarrow\mathcal{B}_t$: relaxation of initial configuration to adaptive reference configuration and the decomposition of the geometric deformation with respect to adaptive reference configuration. }\label{fig:decomposition_map}
\end{figure}
We consider that the adaptive reference configuration $\mathcal{B}_y$ relaxes from the initial reference configuration $\mathcal{B}_0$ to the current deformed configuration $\mathcal{B}_t$; see Fig.~\ref{fig:decomposition_map}. The relaxed reference coordinate is described by the adaptive reference map $\mathbf{y}=\mathbf{Y}(\mathbf{x},t)\in \mathcal{B}_y$, whose dynamics is given by
\begin{equation}\label{eq:y_evol2}
\frac{d\mathbf{y}}{dt}=\frac{\partial \mathbf{Y}}{\partial t}+\mathbf{v}\cdot\nabla \mathbf{Y}=\beta(\mathbf R^T\mathbf x-\mathbf{y}),
\end{equation}
where $\mathbf{y}(\mathbf{x},0)=\mathbf{x}$, $\mathbf{R}$ is the rotational part of the deformation gradient $\mathbf{F}$, and $\beta$ is the rate of rearrangement that measures how fast the reference configuration of the tissue adapts to the current configuration. From the polar decomposition of $\mathbf{F}=\mathbf{R}\mathbf{U}$, where 
$\mathbf{U}=\sqrt{\mathbf{F}^T\mathbf{F}}=\sqrt{\mathbf{C}}$ is the symmetric positive-definite stretch tensor in the material coordinate,  $\mathbf{R}=\mathbf{F}\mathbf{U}^{-1}$ can be uniquely defined as the local rotation (orthogonal transformation $\mathbf{R}^T=\mathbf{R}^{-1}$) that transforms differential elements from the tangent space in the material (curvilinear) coordinate into the tangent space in the current (curvilinear) coordinate. The combination $\mathbf{R}^T{\mathbf{x}}$ is needed to ensure frame invariance of the tensorial evolution equations below under the rotation ${\mathbf{x}}^{+}=\mathbf{Q}(t){\mathbf{x}}$.

With $\beta\neq0$, the adaptive reference configuration $\mathcal{B}_y$ is a new configuration with respect to which we can define the effective deformation gradient $\mathbf{F}=\partial \mathbf{x}/\partial \mathbf{y}=\nabla \mathbf{Y}^{-1}$. The dynamics of $\mathbf{F}$ and its determinant $J$ is consequently affected by $\beta$.
Following similar derivation process in main text with the dynamics of adaptive reference map Eq.~(\ref{eq:y_evol2}), we can get the dynamics of the relaxed deformation gradient 
\begin{equation}\label{eq:F_evol2}
\frac{d (\nabla\mathbf Y^{-1})}{d{{t}}}= \nabla\mathbf{v}\nabla\mathbf{Y}^{-1}-\beta\nabla\mathbf{Y}^{-1}(\mathbf{U}-\mathbf{I})-\beta\nabla\mathbf{Y}^{-1}(\mathbf{x}\cdot\nabla\mathbf{R})\nabla\mathbf{Y}^{-1}.
\end{equation}

Notice that this equation is invariant under the rotation ${\mathbf{x}}^{+}=\mathbf{Q}(t){\mathbf{x}}$ because  $({d (\nabla\mathbf Y^{-1})}/{d{{t}}}-\nabla\mathbf{v}\nabla\mathbf{Y}^{-1})^+=\mathbf{Q}({d (\nabla\mathbf Y^{-1})}/{d{{t}}}-\nabla\mathbf{v}\nabla\mathbf{Y}^{-1})$, $[\nabla\mathbf{Y}^{-1}(\mathbf{U}-\mathbf{I})]^+=[\nabla\mathbf{Y}^{-1}]^+(\mathbf{U}-\mathbf{I})=\mathbf{Q}[\nabla\mathbf{Y}^{-1}](\mathbf{U}-\mathbf{I})=\mathbf{Q}[\nabla\mathbf{Y}^{-1}(\mathbf{U}-\mathbf{I})]$, and $[\nabla\mathbf{Y}^{-1}(\mathbf{x}\cdot\nabla\mathbf{R})\nabla\mathbf{Y}^{-1}]^+=[\nabla\mathbf{Y}^{-1}]^+(\mathbf{x}^+\cdot\nabla_{\mathbf{x}^+}\mathbf{R}^+)[\nabla\mathbf{Y}^{-1}]^+=\mathbf{Q}[\nabla\mathbf{Y}^{-1}](\mathbf{x}\cdot\nabla_{\mathbf{x}}\mathbf{R})[\nabla\mathbf{Y}^{-1}]$, where we have omitted the $\beta$ for simplicity. The first term with $\beta$ demonstrates the adaption of the deformation gradient weighted by the local {\it Biot strain tensor} $(\mathbf{U}-\mathbf{I})$. The second term shows the effect from the rotation $\nabla \mathbf R$ on the dynamics of $\nabla\mathbf Y^{-1}$.

Then the dynamics of relaxed geometric volume variation tracked by the adaptive reference map is  $J=\det{(\nabla\mathbf Y)^{-1}}$: 
\begin{equation}\label{eq:J_evol2}
\frac{dJ}{dt} = J\big[\nabla\cdot\mathbf{v} +{\beta}\Tr(\mathbf{I}-\mathbf{U})-\beta\Tr(\nabla\mathbf{Y}^{-1}(\mathbf{x}\cdot\nabla\mathbf{R}))\big].
\end{equation}

We assume that tissue rearrangement does not interfere the dynamics of tissue density 
\begin{equation}\label{eq:rho_evol1}
\frac{d\rho}{dt} =\frac{\partial \rho}{\partial t}+\mathbf{v}\cdot\nabla\rho= \rho(\gamma - \nabla\cdot \mathbf{v}).
\end{equation}
Combining the relation $\rho J=\rho_0 J_g$ with \cref{eq:rho_evol1,eq:J_evol2}, we obtain the dynamics of $J_g$ with rearrangement
\begin{equation}\label{eq:Jg_evol2}
\frac{dJ_g}{dt} = J_g\big[\gamma +{\beta}\Tr(\mathbf{I}-\mathbf{U})-\beta\Tr(\nabla\mathbf{Y}^{-1}(\mathbf{x}\cdot\nabla\mathbf{R}))\big].
\end{equation}
Assuming that the growth stretch tensor $\mathbf{F}_g$ is still isotropic with tissue rearrangement, the elastic deformation tensor is $\mathbf{F}_e=J_g^{-1/d}\nabla\mathbf{Y}^{-1}$. Thus, the dynamics of the elastic deformation tensor is 
\begin{equation} \label{eq:F_e_evol2}
\frac{d \mathbf{F}_e}{dt}  = (\nabla\mathbf{v}-\boldsymbol{\Gamma}_{iso})\mathbf{F}_e-\beta[\mathbf{R}\left(\mathbf{U}-\mathbf{I}\right)\mathbf{R}^T-\frac{1}{d}\Tr(\mathbf{U}-\mathbf{I})\mathbf{I}]\mathbf{F}_e-\beta[\nabla\mathbf{Y}^{-1}(\mathbf{x}\cdot\nabla\mathbf{R})-\frac{1}{d}\Tr(\nabla\mathbf{Y}^{-1}(\mathbf{x}\cdot\nabla\mathbf{R}))]\mathbf{F}_e.
\end{equation}
The second term $-\beta[\mathbf{R}\left(\mathbf{U}-\mathbf{I}\right)\mathbf{R}^T-\frac{1}{d}\Tr(\mathbf{U}-\mathbf{I})\mathbf{I}] $ contributes to an anisotropic effect on the update of $\mathbf{F}_e$ where $[\mathbf{R}\left(\mathbf{U}-\mathbf{I}\right)\mathbf{R}^T-\frac{1}{d}\Tr(\mathbf{U}-\mathbf{I})\mathbf{I}]$ is traceless. Then the dynamics of the elastic volume variation $J_e=\det\mathbf{F}_e$ is  \begin{equation}\label{eq:Je_evol1}
\frac{dJ_e}{dt}=J_e(\nabla\cdot\mathbf{v}-\gamma).
\end{equation}
One can check that \cref{eq:J_evol2,eq:rho_evol1,eq:Jg_evol2,eq:F_e_evol2,eq:Je_evol1} are all invariant under the rotation ${\mathbf{x}}^{+}=\mathbf{Q}(t){\mathbf{x}}$.

Considering the total strain energy $E(t)=\int_{\Omega_t} J_e^{-1}W(\mathbf{F}_e) dV_t$, the rate of change of total strain energy is given by the dynamics \cref{eq:F_e_evol2,eq:Je_evol1}
\begin{align}
\frac{dE}{dt}=&\int_{\Omega_t} \Big(\boldsymbol{\sigma}:(\nabla \mathbf{v}-\boldsymbol{\Gamma}_{iso})+J_e^{-1}W\gamma\Big) dV_t \nonumber\\
&-\beta \int_{\Omega_t} \boldsymbol{\sigma}_s:[\mathbf{R}\left(\mathbf{U}-\mathbf{I}\right)\mathbf{R}^T-\frac{1}{d}\Tr(\mathbf{U}-\mathbf{I})\mathbf{I}] dV_t\nonumber\\
&-\beta \int_{\Omega_t} \boldsymbol{\sigma}_s:[\nabla\mathbf{Y}^{-1}(\mathbf{x}\cdot\nabla\mathbf{R})-\frac{1}{d}\Tr(\nabla\mathbf{Y}^{-1}(\mathbf{x}\cdot\nabla\mathbf{R}))] dV_t, \label{eq:Energy_rate}
\end{align}
where the stress is given by the constitutive law
\begin{align} 
\boldsymbol\sigma &= \mu J_e^{-\frac{2+d}{d}}\Big(\mathbf{F}_e\mathbf{F}_e^{\text{T}}-\frac{I_1}{d}\mathbf{I}\Big)+K(J_e-1)\mathbf{I}, \nonumber\\ 
&= \mu J_g J^{-\frac{2+d}{d}}\Big(\nabla \mathbf{Y}^{-1}\nabla\mathbf{Y}^{-\text{T}}-\frac{\tilde{I}_1}{d}\mathbf{I}\Big)+K(J_g^{-1}J-1)\mathbf{I} \label{eq:constitutive_law}
\end{align}
where  $\tilde{I}_1=\Tr(\mathbf{C})$ and $I_1=\Tr(\mathbf{C}_e)=\Tr(\mathbf{F}_e^{\text{T}}\mathbf{F}_e)$. The stress can be decomposed as $\boldsymbol{\sigma}=\boldsymbol{\sigma}_s+\boldsymbol{\sigma}_p$, the sum of the deviatoric part $\boldsymbol{\sigma}_s$ (proportional to $\mu$) and the isotropic part $\boldsymbol{\sigma}_p$ (proportional to $K$). 
The first integral in \cref{eq:Energy_rate} represents the energy change due to active growth. The second and the third integrals represent the energy change due to the rearrangement rate $\beta$. Notice $\boldsymbol\sigma$ is replaced by $\boldsymbol\sigma_s$ since $\boldsymbol\sigma_p:[\mathbf{R}\left(\mathbf{U}-\mathbf{I}\right)\mathbf{R}^T-\frac{1}{d}\Tr(\mathbf{U}-\mathbf{I})\mathbf{I}]$ and $\boldsymbol\sigma_p:[\nabla\mathbf{Y}^{-1}(\mathbf{x}\cdot\nabla\mathbf{R})-\frac{1}{d}\Tr(\nabla\mathbf{Y}^{-1}(\mathbf{x}\cdot\nabla\mathbf{R}))]$ both vanish.  Using \cref{eq:constitutive_law}, one can check that $\boldsymbol{\sigma}_s:[\mathbf{R}\left(\mathbf{U}-\mathbf{I}\right)\mathbf{R}^T-\frac{1}{d}\Tr(\mathbf{U}-\mathbf{I})\mathbf{I}] $ is nonnegative and only vanish when $\mathbf{U}=\lambda\mathbf{I}$ is isotropic or $\boldsymbol{\sigma}_s=\mathbf{0}$. This means that the second integral with $\beta>0$ dissipates the stored strain energy only when there are both shear stress and shear deformation. Thus, the adaptive reference map results in physically intuitive properties in the dynamics of the total strain energy in terms related to the Biot strain $\mathbf{U}-\mathbf{I}$. However, the last term involving $\mathbf{x}\cdot\nabla\mathbf{R}$ is not straightforward to interpret geometrically or physically. Thus, we limit our attention to symmetric cases in this work where $\mathbf{x}\cdot\nabla\mathbf{R}=\mathbf{0}$.

\section{Growth of Compressible Tissue in a 2D Disc}
We have also studied the growth of tissue in a 2D disc with fixed height (where there is no deformation in the direction of its thickness). As in Sec. 5 in the main text, we first give the analytic equilibrium solutions for the linearized system. Then we show some numerical simulation results for the nonlinear system. 
\subsection{Mechanical Equilibrium in a 2D Disc}
Writing the linearized system in the approximation of small deformation in the polar coordinates $(r,\theta)$, we obtain
\begin{equation*}
\begin{dcases}
&\frac{\partial u}{\partial t}= v - \beta u, \\
&\frac{\partial g}{\partial t} = \gamma - \beta\Big(\frac{\partial u}{\partial r}+\frac{u}{r}\Big), \\
&\frac{\partial \sigma_{rr}}{\partial r}+\frac{1}{r}(\sigma_{rr}-\sigma_{\theta\theta})=0,\\
& \sigma_{rr}= \mu \Big(\frac{\partial u}{\partial r} - \frac{u}{r}\Big) +K\Big(\frac{\partial u}{\partial r} + \frac{u}{r}-g\Big),\\
& \sigma_{\theta\theta}= \mu \Big(\frac{u}{r}-\frac{\partial u}{\partial r}\Big) +K\Big(\frac{\partial u}{\partial r} + \frac{u}{r}-g\Big),
\end{dcases}
\end{equation*}
subject to deformation-free initial conditions and free moving boundary conditions
\begin{equation*}
\begin{dcases}
u(r,0)=0, \quad g(r,0) = 0,\\
v(0,t)=0,\\
\sigma_{rr}(R(t),t)=0,\quad 
\frac{dR}{dt} = v(R,t).
\end{dcases}
\end{equation*}

When the growth rate $\gamma$ is partially positive and partially negative within the tissue, there exists an equilibrium tissue size $R_{eq}$ such that
\begin{equation}\label{eq:Re_linear}
G(R_{eq})=0, \quad \text{where}\quad G(r) = \int_0^{r} \gamma(\xi) \xi d\xi.
\end{equation}
We can also obtain the solutions in mechanical equilibrium:
\begin{equation}\label{eq:v_linear}
\begin{dcases}
&v= \frac{G(r)}{r} \quad \text{and} \quad
g =\Big(1+\frac{\mu}{ K}\Big)\frac{\gamma(r) }{\beta}, \\
&\sigma_{rr} = -\frac{2\mu G(r)}{\beta r^2}, \\ 
&\sigma_{\theta\theta} = \frac{2\mu}{\beta}\Big(\frac{G(r)}{r^2}-\gamma(r)\Big),\\
&J_e = 1-\frac{\mu }{K}\frac{\gamma(r)}{\beta}.
\end{dcases}
\end{equation}
Similar as in the 3D case, the stresses can be dissipated with large $\beta/\mu$ and the variation in $J_e$ (and also tissue density $\rho=\rho_0/J_e$) is modulated by relative compressibility $K/\mu$ and tissue fluidity relative to growth $\beta/\gamma$. 

\subsection{Nonlinear Simulations}\label{sec:simulation_results}
We perform simulations for the nonlinear system of tissue in a 2D disc
\begin{equation}
\begin{dcases}
&\frac{\partial u}{\partial t} + v\frac{\partial u}{\partial r}= v-\beta u, \\ 
&\frac{\partial J_g}{\partial t} + v\frac{\partial J_g}{\partial r}= J_g\Big[\Gamma + \beta\Big(2-(\frac{\partial y}{\partial r})^{-1}-(\frac{y}{r})^{-1}\Big)\Big], \\ \nonumber
&\frac{\partial \sigma_{rr}}{\partial r}+\frac{1}{r}(\sigma_{rr}-\sigma_{\theta\theta})=0,\\
&\sigma_{rr}= \frac{1}{2}\mu J_g\Big[(\frac{y}{r})^{2}-(\frac{\partial y}{\partial r})^{2}\Big]+K\Big(J_g^{-1}(\frac{\partial y}{\partial r})^{-1}(\frac{y}{r})^{-1}-1\Big), \\ \nonumber
&\sigma_{\theta\theta}= \frac{1}{2}\mu J_g\Big[(\frac{\partial y}{\partial r})^{2}-(\frac{y}{r})^{2}\Big]+K\Big(J_g^{-1}(\frac{\partial y}{\partial r})^{-1}(\frac{y}{r})^{-1}-1\Big),
\end{dcases}
\end{equation}
subject to the deformation-free initial condition and free moving boundary conditions
\begin{equation} 
\begin{dcases}
y(r,0)=r, \quad J_g(r,0) = 1,\\
v(0,t)=0,\\
\sigma_{rr}(R,t)=0,\quad
\frac{dR}{dt} = v(R,t).
\end{dcases}
\end{equation}

\begin{figure}[!h]
	\centering
	\includegraphics[width=0.5\textwidth]{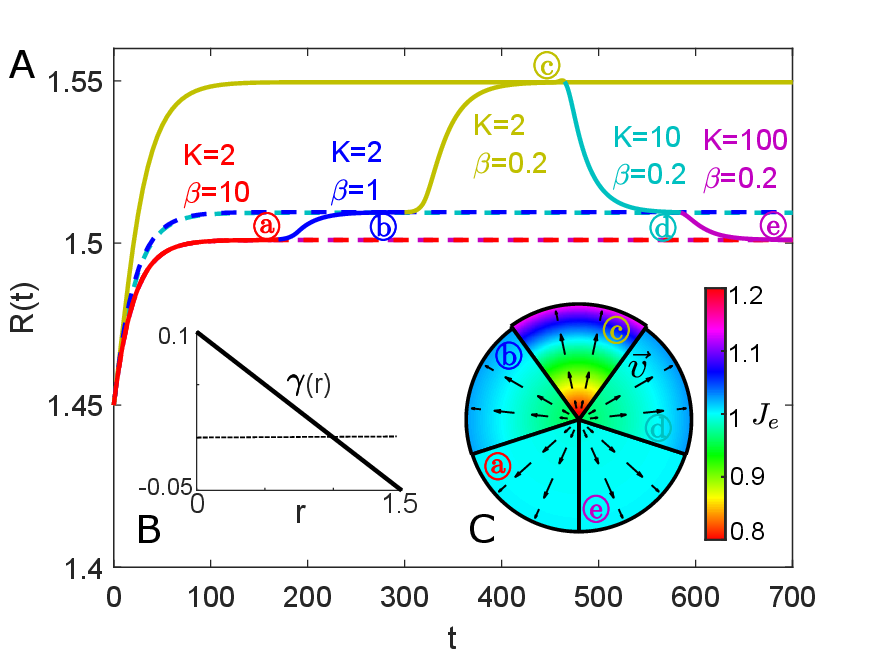}
	\caption{(A) Evolution of tissue size with dynamical modulation of $\beta$ and $K$ (solid line) and with fixed sets of $\beta$ and $K$ (dashed lines) for prescribed $\gamma$. a-e indicate the equilibrium states for each set of parameters $\beta$ and $K$. (B) The spatial profile of  growth rate $\gamma=0.1(1-r)$, where $\gamma>0$ (above dashed line) induces growth and $\gamma<0$ (below dashed line) induces death. (C) The distributions of the inverse of cell density $J_e$ and the tissue flow velocity $v$, at equilibria a-e in (A). The color map shows the values of $J_e$. The arrows indicate the direction and dimensionless magnitude of the tissue velocity. }\label{fig:regulation}
\end{figure}

\begin{figure}[h]
	\centering
	\subfloat[]{\includegraphics[width=0.4\textwidth]{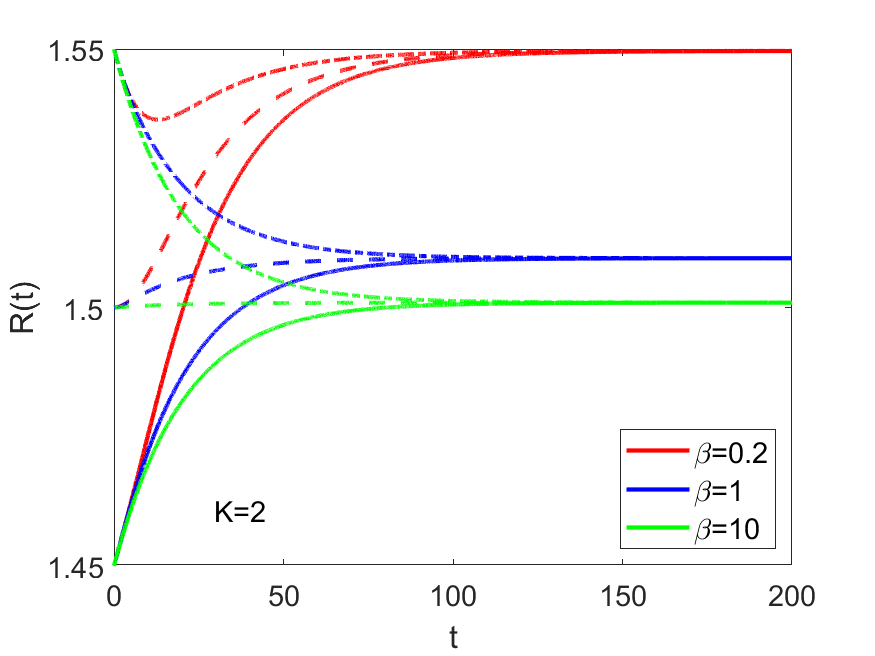}}
	\subfloat[]{\includegraphics[width=0.4\textwidth]{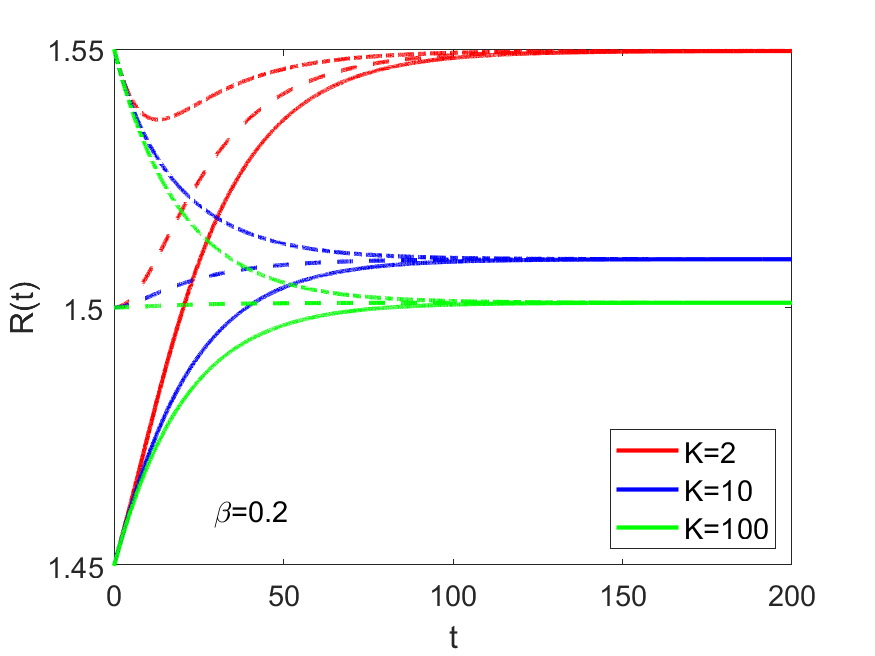}}
	\caption{Evolution of tissue size $R(t)$ with different initial sizes for (a) $K=2$ and $\beta=0.2,1,10$ and (b) $\beta=0.2$ and $K=2,10,100$. Solid lines starts with initial size $R=1.45$, dash lines with $R=1.5$ and dash-dot lines with $R=1.55$. The equilibrium size is independent on the initial size and only dependent on the tissue properties $K$ and $\beta$.  } \label{fig:R_initial}
\end{figure}

\textbf{Nonlinear Effect on Equilibrium Tumor Size.} 
We first simulate the growth of a tissue disc with prescribed growth rate $\gamma=0.1(1-r)$ (Fig.~\ref{fig:regulation} B) and initial radius $R_0=1.45$. The simulation results for the evolution of $R(t)$ for different $K$ and $\beta$ are shown in Fig.~\ref{fig:regulation} A. The inverse of density $J_e=\rho_0/\rho$ (with $\rho_0=1$) in equilibrium for different $K$ and $\beta$ is shown in Fig.~\ref{fig:regulation} C by each sector. The evolution of $R(t)$ with different initial sizes is shown in Fig.~\ref{fig:R_initial}. The equilibrium size is independent on the initial size and only dependent on the tissue properties $K$ and $\beta$.

The equilibrium tissue size $R_{eq}$ for tissue disc satisfies the condition
\begin{eqnarray}\label{eq:equil_size_2d}
G(R_{eq})=\int_0^{R_{eq}} \gamma r dr = -\int_0^{R_{eq}} \frac{1}{J_e} \Big(v\frac{\partial J_e}{\partial r}\Big) r dr,
\end{eqnarray} 
where $\gamma=\gamma(r)$ for the prescribed growth rate and $\gamma=\gamma_R(r)$ for the growth rate coupled with chemical growth factor within the tissue disc of size $R$. This condition demonstrates the coupling of tissue size growth with tissue mechanics.  

\begin{figure}[!h]
	\centering
	\includegraphics[width=0.6\textwidth]{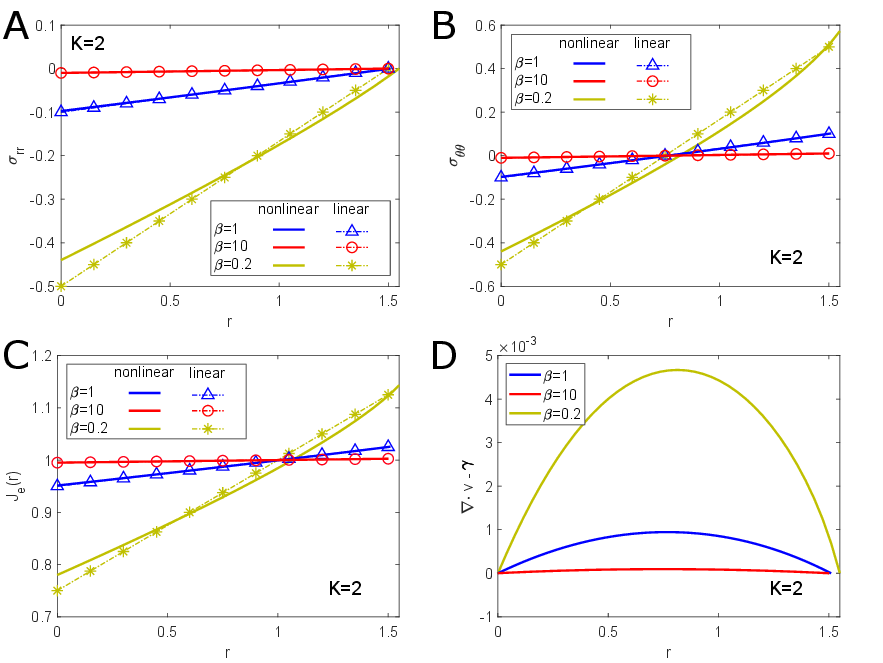}
	\caption{Effect of tissue rearrangement on the mechanical state of tissue at equilibrium for prescribed $\gamma$. (A and B) Radial and circumferential stresses. (C) The elastic stretch $J_e$. Cells are compressed and concentrated near the center and stretched near the boundary. The nonlinear solutions (solid lines) agrees with the linear solutions (dashed lines with markers) when the tissue rearrangement is much faster than tissue growth ($\beta=1, 10$). For slower rearrangement ($\beta=0.2$), the nonlinear solutions start to deviate from linear solutions. (D) Compressibility of tissue flow. For incompressible flow, $\nabla\cdot v-\gamma=0$. }\label{fig:beta_regulation}
\end{figure}

\begin{figure}[!h]
	\centering
	\subfloat[]{\includegraphics[width=0.33\textwidth]{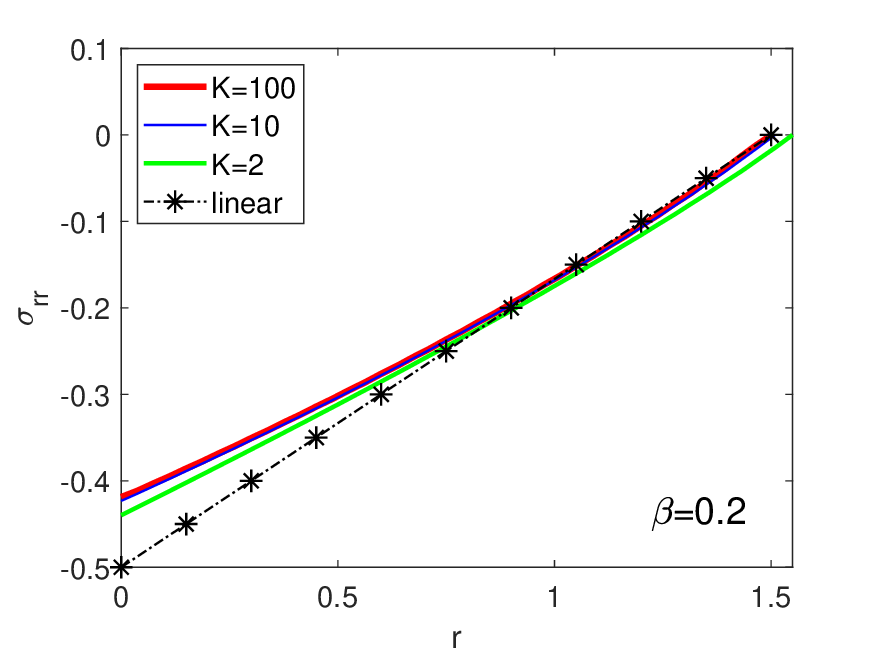}}
	\subfloat[]{\includegraphics[width=0.33\textwidth]{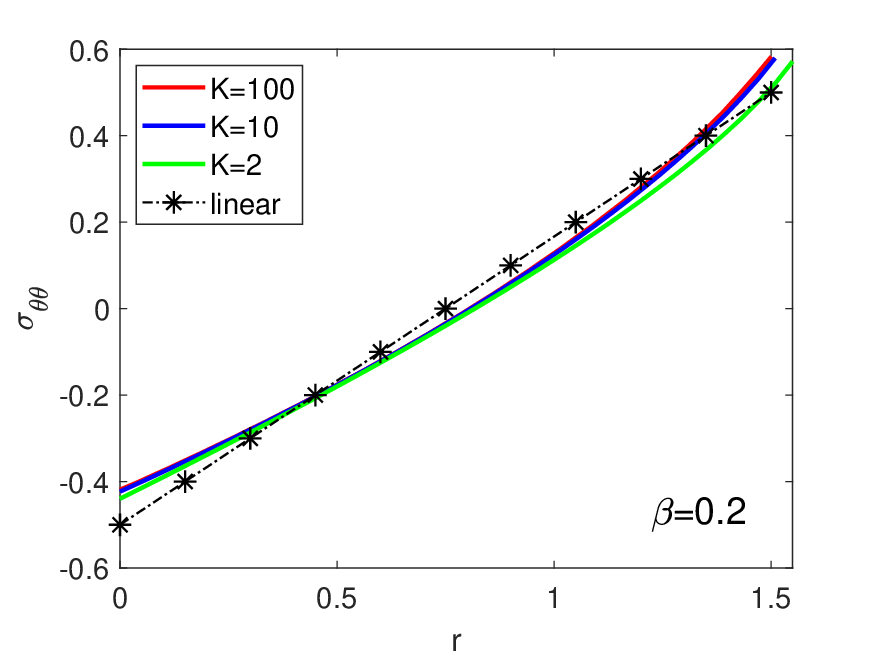}}
	\caption{Effect of incompressibility $K$ on (a) radial stress $\sigma_{rr}$ and (b) circumferential stress $\sigma_{\theta\theta}$ with $\beta=0.2$ for prescribed $\gamma$. } \label{fig:K_regulation}
\end{figure}

\textbf{Regulation of Mechanical Stress and Tissue Density.} The mechanical states including stresses $\sigma_{rr}$, $\sigma_{\theta\theta}$ and elastic volume variation $J_e$ in equilibrium for different $\beta$ are shown in Fig.~\ref{fig:beta_regulation}. They can be dissipated by fast rearrangement (large $\beta$), just as in 3D case. In particular, the incompressibility condition $\nabla\cdot \mathbf{v}-\gamma$ is achieved for $\beta=10$ (Fig.~\ref{fig:beta_regulation} D), as predicted in linear solutions. The stresses are scarcely altered by the change of $K$, as shown in Fig.~\ref{fig:K_regulation}.

\begin{figure}[!h]
	\centering
	\includegraphics[width=0.49\textwidth]{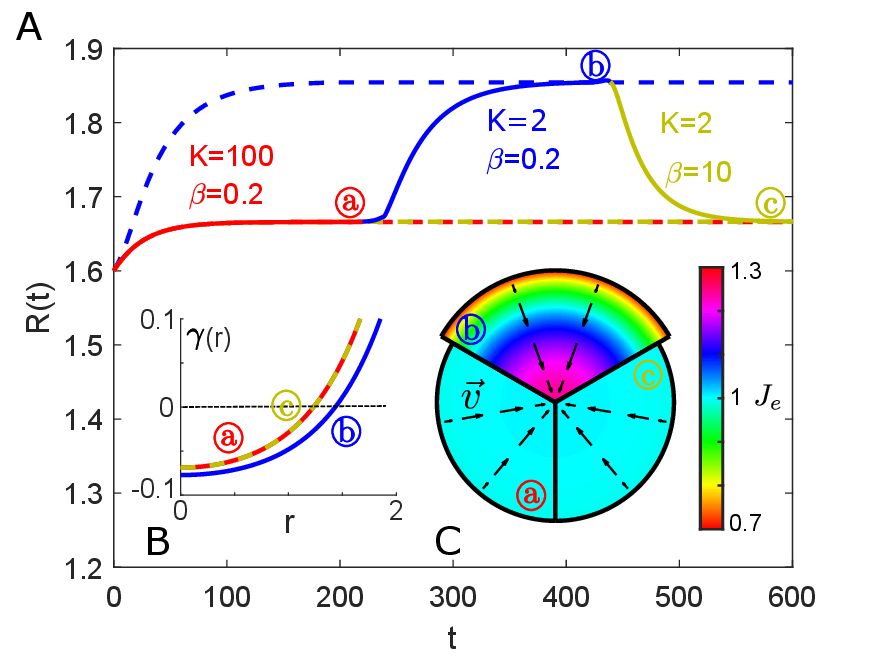}
	\includegraphics[width=0.49\textwidth]{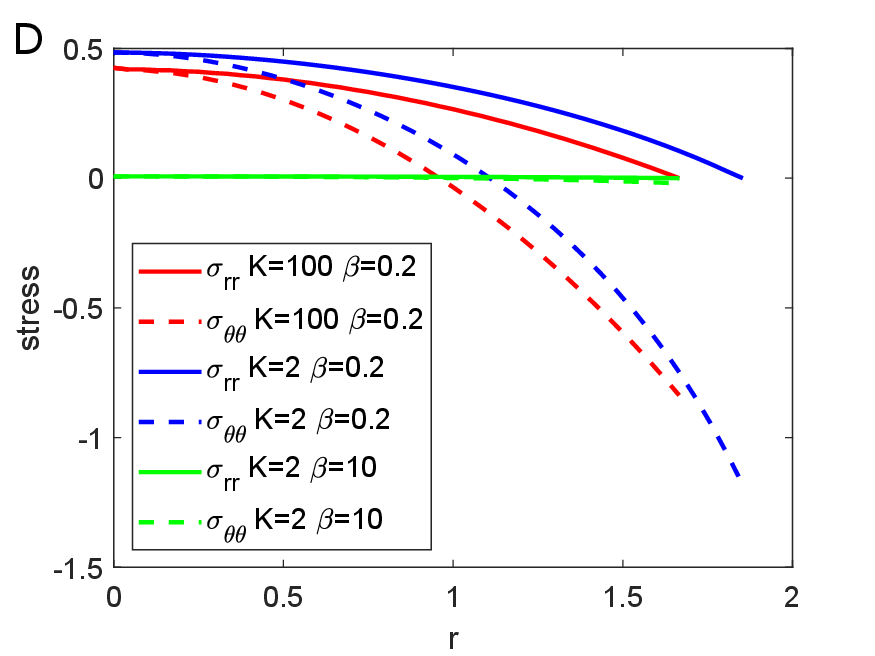}
	\caption{(A) Evolution of tissue size with dynamical modulation of $\beta$ and $K$ (solid line) and with fixed sets of $\beta$ and $K$ (dashed lines) for $\gamma$ coupled with growth factor. a-c indicate the equilibria for each set of parameters $\beta$ and $K$. (B) The spatial profile of $\gamma$, coupled with the reaction-diffusion process of growth factor (with $L=0.5$, $\lambda_p=0.2$ and $\lambda_a=0.1$). The profiles of $\gamma$ for the equilibrium a and c are identical due to the same equilibrium tissue size $R(t)$. (C) The inverse of cell density $J_e$ and the tissue flow velocity $v$, at equilibria a-c in (A). The color map shows the values of $J_e$. The arrows indicate the direction and dimensionless magnitude of the tissue velocity. (D) Modulation of stresses by $K$ and $\beta$.}\label{fig:coupling}
\end{figure}

\textbf{Growth Rate Coupled with Chemical Signals.} We then simulate the growth of a tissue disc when the growth rate $\gamma_R(r)$ is coupled with the concentration of growth factor $c_R(r)$
\begin{equation}
\label{eq:gammac}
\gamma_R(r) =  c_R(r) \lambda_{p}- \lambda_{a},
\end{equation}
where $c_R\lambda_{p}$ and $\lambda_{a}$ are the local rates of cell proliferation and apoptosis, respectively. The concentration of growth factor within a tissue disc of size $R(t)$ follows the reaction-diffusion process 
\begin{equation}\label{eq:diffusion}
\begin{dcases}
&\frac{1}{\lambda}\frac{\partial c(r)}{\partial t} = L^2\frac{1}{r}\frac{\partial}{\partial r}\Big(r\frac{\partial c}{\partial r}\Big)- c,  \quad  \text{for $0< r< R(t)$,}\\
& c(R(t),t)=1, \quad \frac{\partial c}{\partial r}(0,t)=0.
\end{dcases}
\end{equation}
Assuming $\lambda\gg \gamma$, we can obtain the quasi-steady-state solution $c_R(r)$ of \cref{eq:diffusion} for tissue radius $R(t)$ at each time point:
\begin{equation}
c_R(r) = I_0(r/L)/I_0(R/L),
\end{equation}
where $I_0(x)$ is the modified Bessel function of first kind. 
We perform simulation for the growth of a tissue disc with $\gamma_R$. The evolution of $R(t)$ for different $K$ and $\beta$ is shown in Fig.~\ref{fig:coupling} A. The growth rate $\gamma_R$ in equilibria is shown in Fig.~\ref{fig:coupling} B and the elastic volume variation $J_e$ in equilibria is shown in Fig.~\ref{fig:coupling} C. The stresses can be dissipated by large $\beta$ and can be moderately reduced by large $K$, as shown in Fig.~\ref{fig:coupling} D.

